\documentclass[%
 reprint,
 amsmath,amssymb,
prb,
]{revtex4-1}

\usepackage{graphicx}
\usepackage{dcolumn}
\usepackage{bm}

\newcommand{\be}{\begin{equation}}
\newcommand{\ee}{\end{equation}}

\newcommand{\bea}{\begin{eqnarray}}
\newcommand{\eea}{\end{eqnarray}}

\newcommand{\nn}{\nonumber}

\begin{document}

\preprint{APS/123-QED}

\title{Scaling laws for band gaps of phosphorene nanoribbons: A tight-binding calculation}

\author{Esmaeil Taghizadeh Sisakht}
\author{Mohammad H. Zare}
\author{Farhad Fazileh}
\email{fazileh@cc.iut.ac.ir}
\affiliation{Department of Physics, Isfahan University of Technology, Isfahan 84156-83111, Iran}

\date{\today}

\begin{abstract}
In this study, we analyze the band structure, the state characterization, and electronic transport of monolayer black phosphorus (phosphorene) zigzag nanoribbons (zPNRs) and armchair nanoribbons (aPNRs), using five-parameter tight-binding (TB) approximation. In zPNRs, the ratio of the two dominant hopping parameters indicates the possibility of a relativistic dispersion relation and the existence of a pair of separate quasi-flat bands at the Fermi level. Moreover, the corresponding states are edge localized if their bands are well separated from the valence and conduction bands. We also investigated the scaling laws of the band gaps versus ribbon widths for the armchair and zigzag phosphorene nanoribbons. In aPNRs, the transverse electric field along the ribbon width enhances the band gap closure by shifting the energy of the valence and conduction band edge states. For zPNRs, a gap occurs at the middle of the relatively degenerate quasi-flat bands; thus, these ribbons are a promising candidate for future field-effect transistors.  
\end{abstract}

\maketitle


\section{\label{introduction}Introduction}
Two-dimensional (2D) structures that are inspired by graphene such as hexagonal boron nitride (BN) and transition metal dichalcognides (TMDs) have attracted considerable attentions owing to their remarkable electronic properties \cite{Novoselov2004,Geim2007,Neto2009,Splendiani2010,Mak2010,Xiao2010,Blase1995,Watanabe2004}. Graphene is known to have novel electronic and mechanical properties such as high carrier mobility; however, its zero band gap limits its performance. As a TMD, molybdenum disulphide (MoS$_2$) has a direct band gap of $\sim$ 1.8 eV \cite{Kuc2011} and a relatively high on/off ratio \cite{Radisavljevic2011}. However, the carrier mobility of MoS$_2$ is much less than that of graphene. These layered structures can be etched or patterned as quasi-one-dimensional (1D) strips referred to as nanoribbons. Graphene nanoribbons (GNRs) and MoS$_2$ nanoribbons are examples of these 1D strips. These 1D nanoribbons can offer better tunability in electronic structures because of quantum confinement and 
edge effects \cite{Son2006,Yang2007,Wang2008}.

\begin{figure}[t]
\centering
\includegraphics[width=0.47\textwidth]{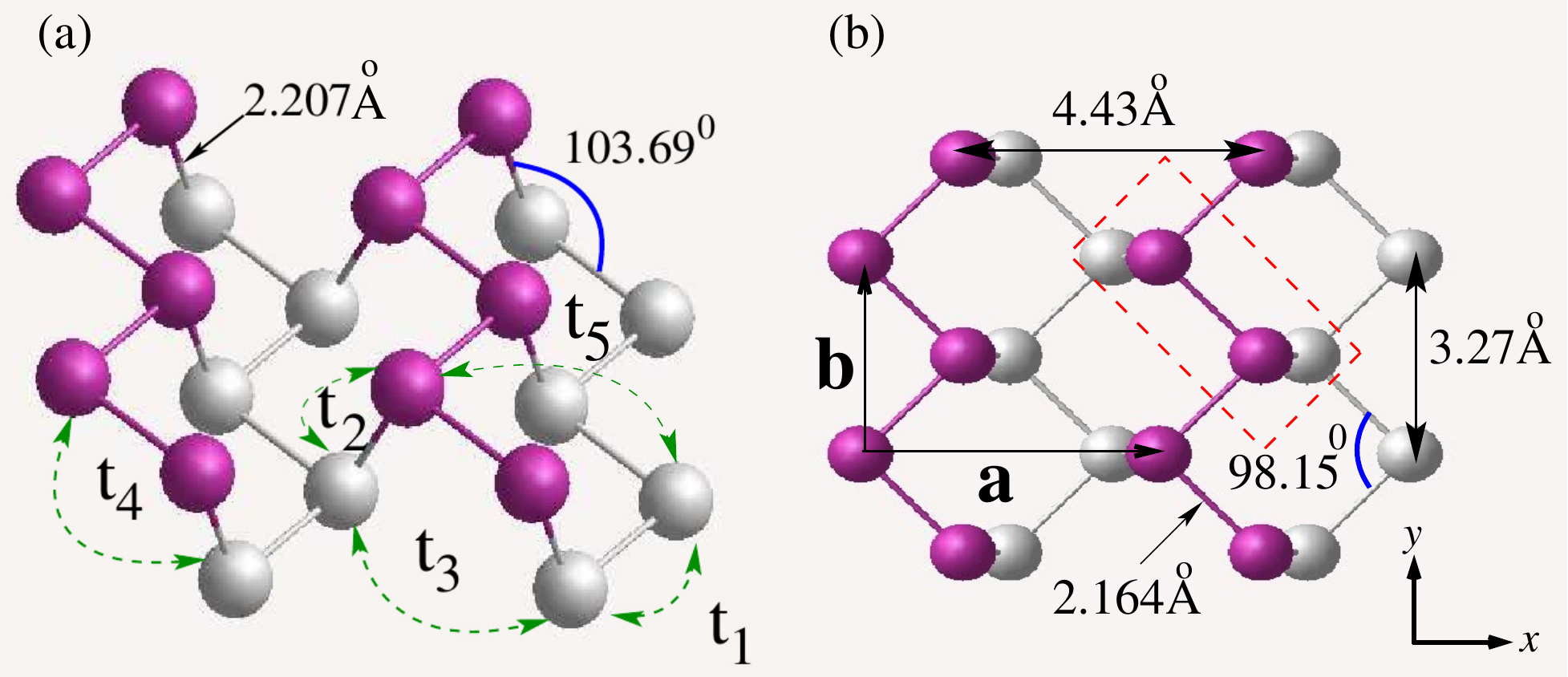}
\caption{(a) Crystal structure and hopping integrals $t_i$ of single layer phosphorene for the TB model. (b) Top view. Note that the dark (gray) balls represent the phosphorus atoms in the upper (lower) layer. The dotted rectangle indicates a primitive unit cell containing four atoms. The parameters for the bond angles and unit cell lengths are taken from \cite{Gomez2014}.}
\label{fig:Lattice}
\end{figure}

Monolayer black phosphorus, referred to as phosphorene, has attracted much attention recently because of its potential applications in nano-electronics, thermo-electronics and opto-electronics \cite{Li2014,Liu2014,Xia2014,Gomez2014,Koenig2014}. Phosphorene has a finite band gap and greater mobility as compared with MoS$_2$. Similar to bulk graphite, black phosphorus is also a layered structure in which the layers are held together by Van der Waals interactions \cite{Morita1986}. Each layer consists of phosphorus atoms that are covalently bonded to three adjacent phosphorus atoms, thus forming a puckered honeycomb structure because of $sp^3$ hybridization, as shown in Fig.~\ref{fig:Lattice}. As can be seen, the phosphorus sites are grouped in two zigzag layers. The upper and lower sites are shown with darker and lighter colors, respectively. Phosphorene has been successfully fabricated in the laboratory by numerous researchers \cite{Li2014,Liu2014,Xia2014,Gomez2014,Koenig2014}. Graphene can be isolated by peeling;
 similarly, phosphorene can also be isolated from black phosphorus via mechanical exfoliation. Phosphorus has a direct band gap of 0.3 eV \cite{Liu2014,Warschauer1963,Narita1983,Maruyama1981}. Phosphorene layers can be mechanically exfoliated from bulk phosphorus, and the band gap of phosphorene thus obtained ranges from 2.0 eV (monolayer) to 0.6 eV (five-layers)~\cite{Liang2014,Tran20141,Qiao2014}. Although phosphorene nanoribbons (PNRs) have not yet been fabricated, experience from graphene and other 2D materials suggests the electronic structure and optical properties of PNRs must be studied for future research on phosphorene-based nanoelectronics. Numerous studies have focused on first-principle calculations \cite{Wei2014,Zhang2014,Lv2014,Gong2014}. Recently, a TB model has been proposed by introducing hopping integrals ($t_i$) over five neighbouring sites \cite{Rudenko2014}, as shown in Fig.~\ref{fig:Lattice}(a).

Our goal is to apply the above mentioned TB model to zigzag and armchair phosphorene nanoribbons to analyze their band structure and quantum conductance and compare the results with other more sophisticated calculations. Thereafter, we examine the effect of transverse electric field on the band structure and quantum conductance of both zigzag and armchair nanoribbons.

In section \ref{ModelHamiltonian}, the TB model is introduced. In section \ref{Bulk}, the band structure and effective masses of the monolayer phosphorene near the gap are presented based on the TB model and it is shown that the dispersion is relativistic along the $y$ direction and the Fermi velocities along this direction are calculated. In section \ref{Numericalresults}, the numerical data for this model is presented for zPNRs and aPNRs, and the emergence of edge states and the gradual emergence of flat bands in zPNRs when $|t_2/t_1|$ ratio is increased is discussed. The scaling behavior of band gap with ribbon width is presented and the obtained results are compared with those of the other methods. In addition, the effect of transverse electric field on the band gap in aPNRs and the transistor effect in zPNRs are investigated.

\section{\label{ModelHamiltonian}Model Hamiltonian}
The TB Hamiltonian recently proposed for this system is given by \cite{Rudenko2014}
\begin{equation}
H = \sum_{i,j} t_{ij}c_i^{\dagger}c_j 
\label{eqn:TBhamiltonian}
\end{equation}
where the summation is over the lattice sites, and $t_{ij}$ are the hopping integrals between the $i$th and $j$th sites. Further, $c_i^{\dagger}$ and $c_j$ represent the creation and annihilation operators of electrons in sites $i$ and $j$, respectively.
These hopping integrals between a site and its neighbours are shown in Fig.~\ref{fig:Lattice}(a). 

The connections in the upper or lower layers in each zigzag chain are represented by $t_1$ hopping integrals, and the connections between a pair of upper and lower zigzag chains are represented by $t_2$ hopping integrals. Further, $t_3$ denotes the hopping integrals between the nearest sites of a pair of zigzag chains in the upper or lower layer, and $t_4$ denotes the hopping integrals between the next nearest neighbor sites of a pair of upper and lower zigzag chains. Finally, $t_5$ is the hopping integrals between two atoms on the upper and lower zigzag chains that are farthest from each other. The specific values of these hopping integrals as suggested in 
~\cite{Rudenko2014} are as follows: $t_1=-1.220$~eV, $t_2=3.665$~eV, $t_3=-0.205$~eV, $t_4=-0.105$~eV, and $t_5=-0.055$~eV. The special characteristic of this model is that the second hopping integral is positive. This implies that the zigzag chains have negative $t_1$ hopping integrals along the chains and positive $t_2$ hopping integrals connecting these chains. For zPNRs, the eigenstates of the transverse modes, which characterize the behavior of the states as edge or bulk states, are along both $t_1$ and $t_2$ connections. The role of this behavior in creation of a relativistic band dispersion along the $\Gamma$-X direction will be discussed in the next section.

\section{\label{Bulk}Monolayer phosphorene}
In this section the band structure and effective masses of the electron and hole states of the bulk monolayer phosphorene is calculated based on the above mentioned TB model and the results are compared with ab-inito calculations. Since each unit cell of a single layer phosphorene contains four phosphorus atoms [Fig.~\ref{fig:Lattice}(b)], initially, a four band model is created. The band dispersion along the two periodic directions of $\Gamma$-X and $\Gamma$-Y are compared and the electron and hole effective masses are compared along the two directions. In the next subsection it is argued that the unit cell for the electronic model only contains two phosphorus atoms resulting in a two band model. Finally, the band gap at $\Gamma$ point is derived as a function of the hopping parameters.

\begin{figure}[t]
\centering
\vspace{20pt}
\includegraphics[width=0.45\textwidth]{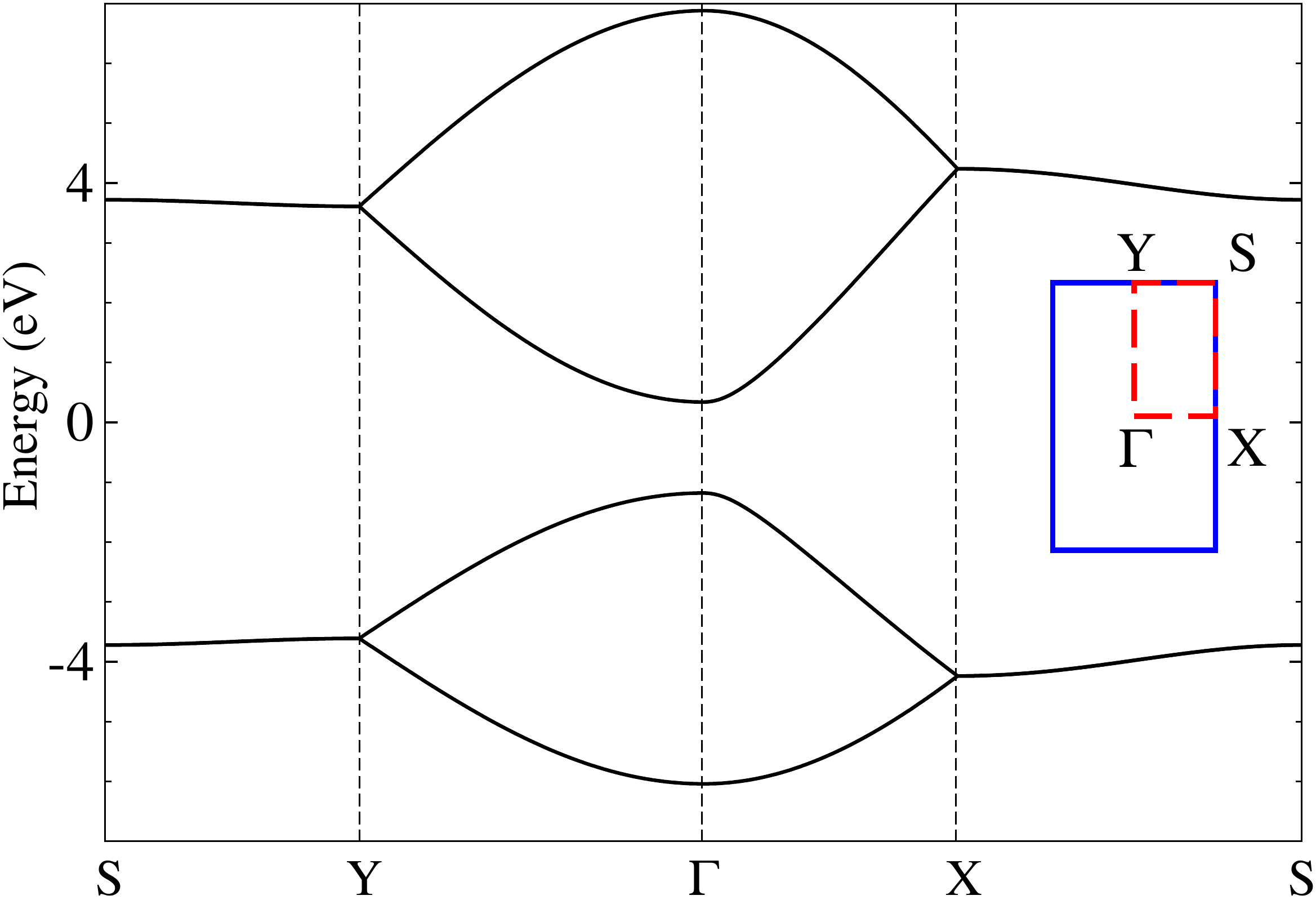}
\caption{Tight-binding energy band structure for bulk phosphorene.}
\label{fig:BEB1}
\end{figure}

\subsection{\label{4band}Four-band tight-binding model}
As shown in Fig.~\ref{fig:Lattice}(b) the unit cell of the monolayer phosphorene is a rectangle containing four phosphorus atoms.   
Fourier transforming, the general Hamiltonian in momentum space is given by:
\be
H=\sum_{\bf k} \psi^{\dagger}_{\bf k}H^{[4]}_{\bf k}\psi_{\bf k}
\label{h4}
\ee
where $\psi^{\dagger}_{\bf k}=(a^{\dagger}_{\bf k}~~b^{\dagger}_{\bf k}~~c^{\dagger}_{\bf k}~~d^{\dagger}_{\bf k})$
and  $H^{[4]}_{\bf k}$ is a $ 4\times 4 $ matrix
\be
H^{[4]}_{\bf k} = \begin{pmatrix} 0 & A_{\bf k}& B_{\bf k} & C_{\bf k}  \\
A^{*}_{\bf k} & 0 & D_{\bf k} & B_{\bf k}  \\ 
B^{*}_{\bf k} & D^{*}_{\bf k} & 0 & A_{\bf k}  \\
C^*_{\bf k} & B^{*}_{\bf k} & A^{*}_{\bf k} & 0  \\
\label{Hk4}
\end{pmatrix}
\ee
whose elements are given by

\bea
A_{\bf k} &=& t_2+t_5e^{-ik_a} \nn \\
B_{\bf k} &=& 4t_4e^{-i(k_a-k_b)/2}\cos(k_a/2)\cos(k_b/2) \nn \\
C_{\bf k} &=& 2e^{ik_b/2}\cos(k_b/2)(t_1e^{-ik_a}+t_3) \nn \\
D_{\bf k} &=& 2e^{ik_b/2}\cos(k_b/2)(t_1+t_3e^{-ik_a}). \nn \\
\label{ElementsHK}
\eea

Here $ k_a={\bf k}\cdot{\bf a}$ and $k_b={\bf k}\cdot{\bf b}$,
 where ${\bf a}=a\hat{\bf x}$ and ${\bf b}=b\hat{\bf y}$ are the primitive translational  
vectors of the structure displayed in Fig.~\ref{fig:Lattice}(b).
Bulk energy bands for the monolayer phosphorene are shown in Fig.~\ref{fig:BEB1}.
The band dispersion is relativistic along the $x$ direction whereas it is nonrelativistic along the $y$ direction.
Considering a relativistic band dispersion, $E=\sqrt{m^2v_F^4+p^2v_F^2}$, along the $\Gamma$-X direction  and a parabolic form along the $\Gamma$-Y direction near the conduction band minimum (CBM) and valence band maximum (VBM) the effective masses and the Fermi velocities are calculated and presented in Table~\ref{tab:table1}.
It can be deduced from Table~\ref{tab:table1} that electrons and holes moving along the zigzag direction are more than six times heavier than those moving along the armchair direction.

\begin{table}
\caption{\label{tab:table1}%
Fermi velocities and effective masses of electron and hole states near the CBM and VBM along the two directions of $\Gamma$-X and $\Gamma$-Y.
}
\begin{ruledtabular}
\begin{tabular}{lcc}
\textrm{Band}&
\textrm{$v_F$ ($\times 10^5~m/s $)}&
\textrm{$m/m_0$}\\
\colrule
$\Gamma$-X (e) & 9.71 & 0.164 \\
$\Gamma$-X (h) & 8.26 & 0.179 \\
$\Gamma$-Y (e) & -- & 0.873 \\
$\Gamma$-Y (h) & -- & 1.175 \\
\end{tabular}
\end{ruledtabular}
\end{table}

\begin{figure} 
\centering
\vspace{20pt}
\includegraphics[width=0.45\textwidth]{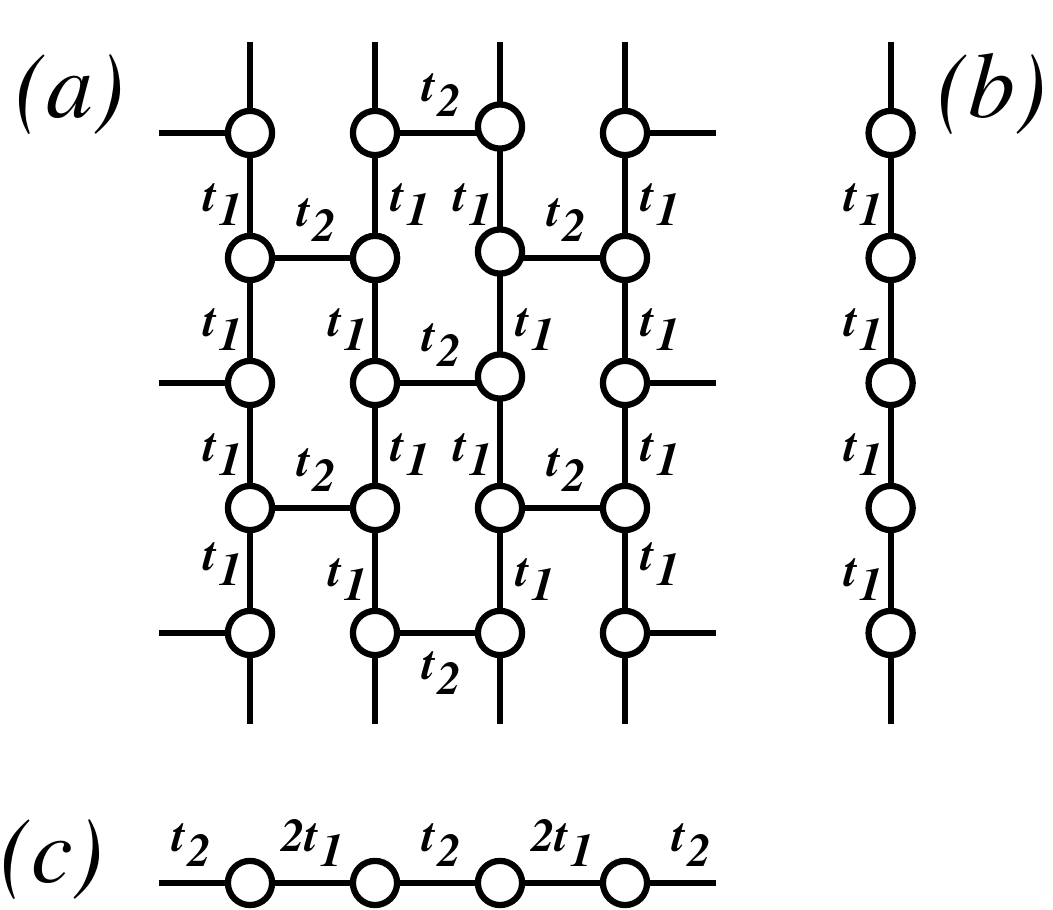}
\caption{(a) Topologically equivalent structure to monolayer phosphorene within two parameter TB model. (b) Equivalent linear chain model along the zigzag direction. (c) Equivalent linear chain model along the armchair direction.}
\label{fig:eq_model}
\end{figure}

There is a simple explanation for the reason why this special combination for the dominant hopping parameters ($t_1=-1.220$~eV and $t_2=3.665$~eV) creates a nearly relativistic dispersion near $\Gamma$ point along the $x$ direction. We introduce a lattice model [Fig.~\ref{fig:eq_model}(a)], which is equivalent to the monolayer phosphorene within the two parameter TB approximation. For this model the dispersion along the $y$ direction for large wavelengths along $x$ ($|k_a|\sim 0$ and no dynamics along the $x$ direction) can be modeled by TB on a linear chain shown in Fig.~\ref{fig:eq_model}(b). Similarly, the dispersion along the $x$ direction for $|k_b|\sim 0$ can be modeled by TB on a chain shown in Fig.~\ref{fig:eq_model}(c). For the linear chain of Fig.~\ref{fig:eq_model}(b) the dispersion would be $2t_1\cos(k_b)$, which near $k_b \simeq 0$ is $-|2t_1|+|t_1|k_b^2$, and it is parabolic. This dispersion gives rise to an effective mass of $m=1.17m_0$, which is consistent with the data in Table~\ref{tab:table1}. The dispersion for the linear chain of Fig.~\ref{fig:eq_model}(c) along the $x$ direction is given by $\pm \sqrt{(2t_1)^2+t_2^2+4t_1t_2\cos(k_a)}$. In terms of the absolute values of the hopping parameters and near the $k_a \simeq 0$, this relation is reduced to $\pm \sqrt{(|2t_1|-|t_2|)^2 + 2|t_1t_2|k_a^2}$. When $|t_2|$ is close to $|2t_1|$, we can ignore the first term under the square root and the dispersion will be linear $\pm \sqrt{2|t_1t_2|}k_a$ and the constant of proportionality gives a Fermi velocity of $2\pi \sqrt{2(1.22\textrm{eV})(3.665\textrm{eV})}\times(4.43\textrm{\AA}/2)/(12400\textrm{eV\AA}) \times c \sim 10^6$~m/s which is consistent with the data of Table~\ref{tab:table1}. For the model of Eq.\ref{eqn:TBhamiltonian}, $|t_2| \simeq 3|t_1|$ which does not give an exactly linear dispersion but it gives a massive relativistic dispersion, and for larger values of $k_a$ it is nearly linear.     

\subsection{\label{2band}Two-band tight-binding model}
In the TB Hamiltonian of Eq.~\ref{eqn:TBhamiltonian}, if we project the positions of the upper and lower zigzag chains on a horizontal plane and keep the previous hopping integrals, the unit cells of the electronic system is reduced to two phosphorus atoms per unit cell. The Fourier transform of the resulting two band model is given by

\be
H=\sum_{\bf k} \phi^{\dagger}_{\bf k}H^{[2]}_{\bf k}\phi_{\bf k}
\label{h4}
\ee 
where $\phi^{\dagger}_{\bf k}=(a^{\dagger}_{\bf k}~~b^{\dagger}_{\bf k})$
and  $H^{[2]}_{\bf k}$ is a $ 2\times 2 $ matrix

\be
H^{[2]}_{\bf k} = \begin{pmatrix} B_{\bf k} e^{i(k_a-k_b)/2} &~~ A_{\bf k}+C_{\bf k}e^{i(k_a-k_b)/2}  \\
A^{*}_{\bf k}+C^{*}_{\bf k}e^{-i(k_a-k_b)/2} &~~ B_{\bf k} e^{i(k_a-k_b)/2}   
\label{Hk2}
\end{pmatrix}
\ee

Diagonalizing the above matrix, the energy spectrum is

\be
E_{\bf k}=|B_{\bf k}|\pm|A_{\bf k}+C_{\bf k}e^{i(k_a-k_b)/2}|
\label{spectrum}
\ee

The band gap in the $\Gamma$ point is
\be
E_{g}=4t_1+2t_2+4t_3+2t_5=1.52~\textrm{eV}.
\ee

\section{\label{Numericalresults}Electronic and transport properties of Phosphorene nanoribbons}
 In the following numerical analysis, the commonly used method for determining the width of graphene nanoribbons~\cite{Son2006} is employed to determine the PNR structures. According to this method, the structure of aPNR is defined by the number of dimmer lines across the ribbon width ($N_a$-aPNRs), whereas that of zPNR is defined by the number of zigzag chains across the ribbon width ($N_z$-zPNRs)~\cite{Tran2014}. To calculate the band structure and eigenstates of the nanoribbons, we obtain the eigenvalues and eigenvectors of the following matrix, which is the crystal Hamiltonian between Bloch sums:
\begin{equation}
M_{\alpha\beta}({\bf k}) = -\sum_{ij}t_{i\alpha;j\beta} e^{i{\bf k}\cdot{\bf R}_{ij}} 
\label{eqn:band1}
\end{equation}
where $i$ and $j$ denote different unit cells, $\alpha$ and $\beta$ denote the basis sites in a unit cell. Further, $\bf{k}$ is the wave vector, and ${\bf R}_{ij}$ represents a bravais lattice vector. Moreover, $t_{i\alpha;j\beta}$ are the hopping integrals between the basis site $\alpha$ of unit cell $i$ and the basis site $\beta$ of unit cell $j$, and will be substituted by the five hopping parameters of the model, accordingly. For nanoribbons, the periodicity is only along the ribbon length; therefore, the number of basis sites in each unit cell is proportional to the ribbon width.

\begin{figure}[b] 
\centering
\includegraphics[width=0.45\textwidth]{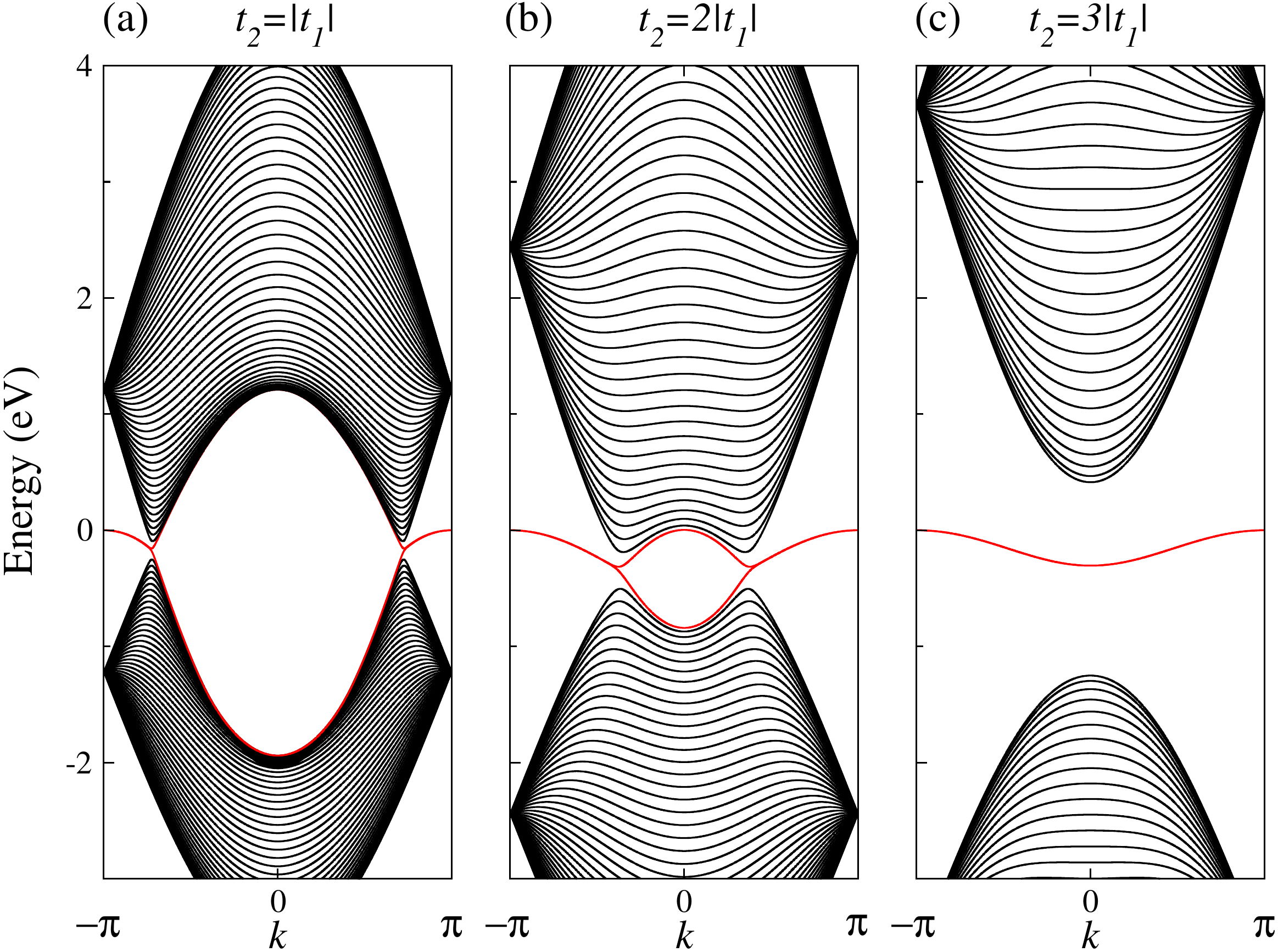}

\vspace{20pt}
\centering
\includegraphics[width=0.48\textwidth]{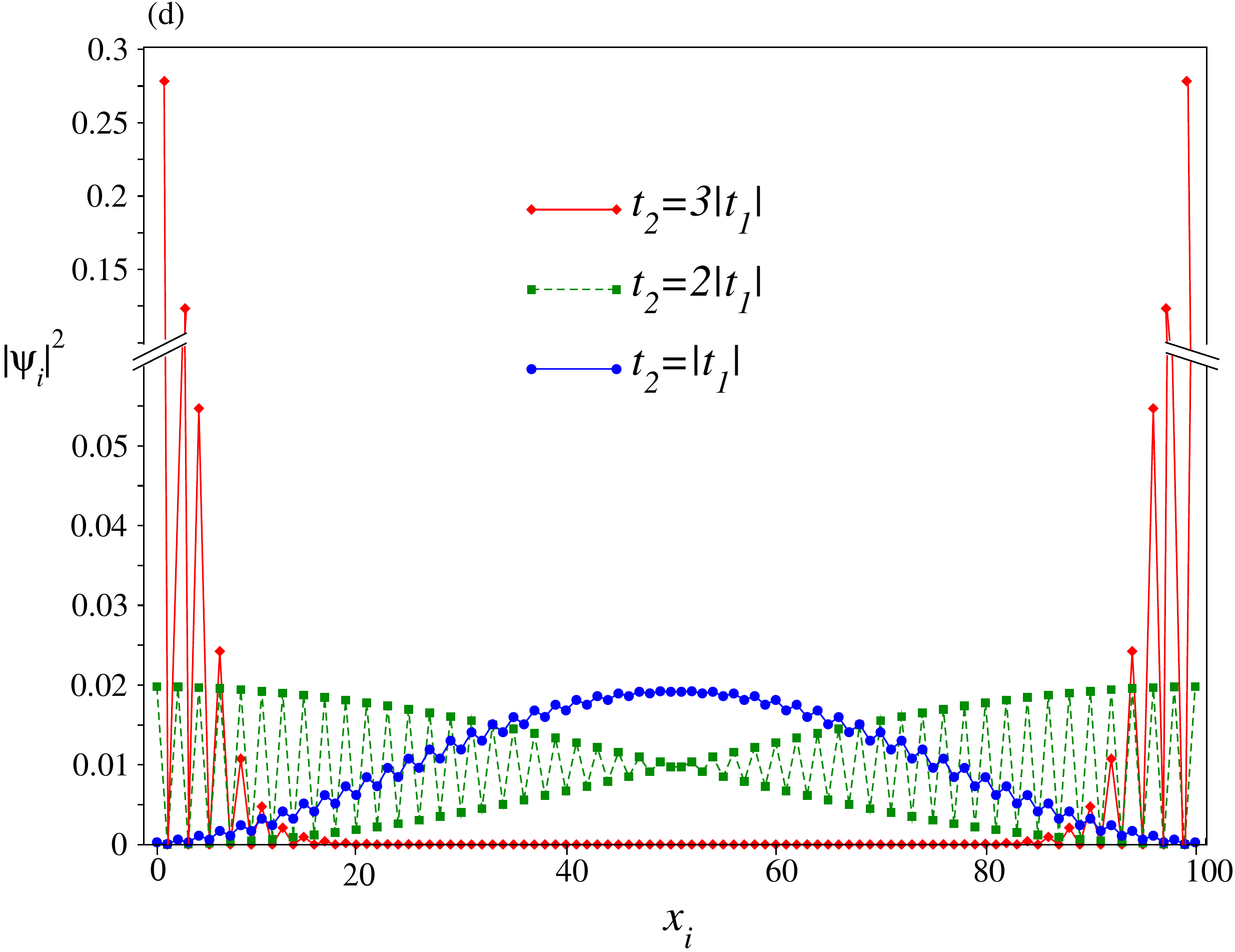}
\caption{Top: Band structure of 100-zPNRs ($w \sim$ 22~nm) for $|t_2/t_1|$ ratio values of (a) $|t_2/t_1|$=1, (b) $|t_2/t_1|$=2, and (c) $|t_2/t_1|$=3 for $t_3=-0.205$, $t_4=-0.105$, and $t_5=-0.055$. Note that the red lines represent the edge bands. Bottom:  The probability amplitude of the upper valence band eigenstate for $k=0$ of a zigzag phosphorene nanoribbon for different ratios of $|t_2/t_1|$. Note that the horizontal axis represents a unit cell in the width of the ribbon.}
\label{fig:t2t1}
\end{figure}

\begin{figure}[t]
\centering
\vspace{20pt}
\includegraphics[width=0.45\textwidth]{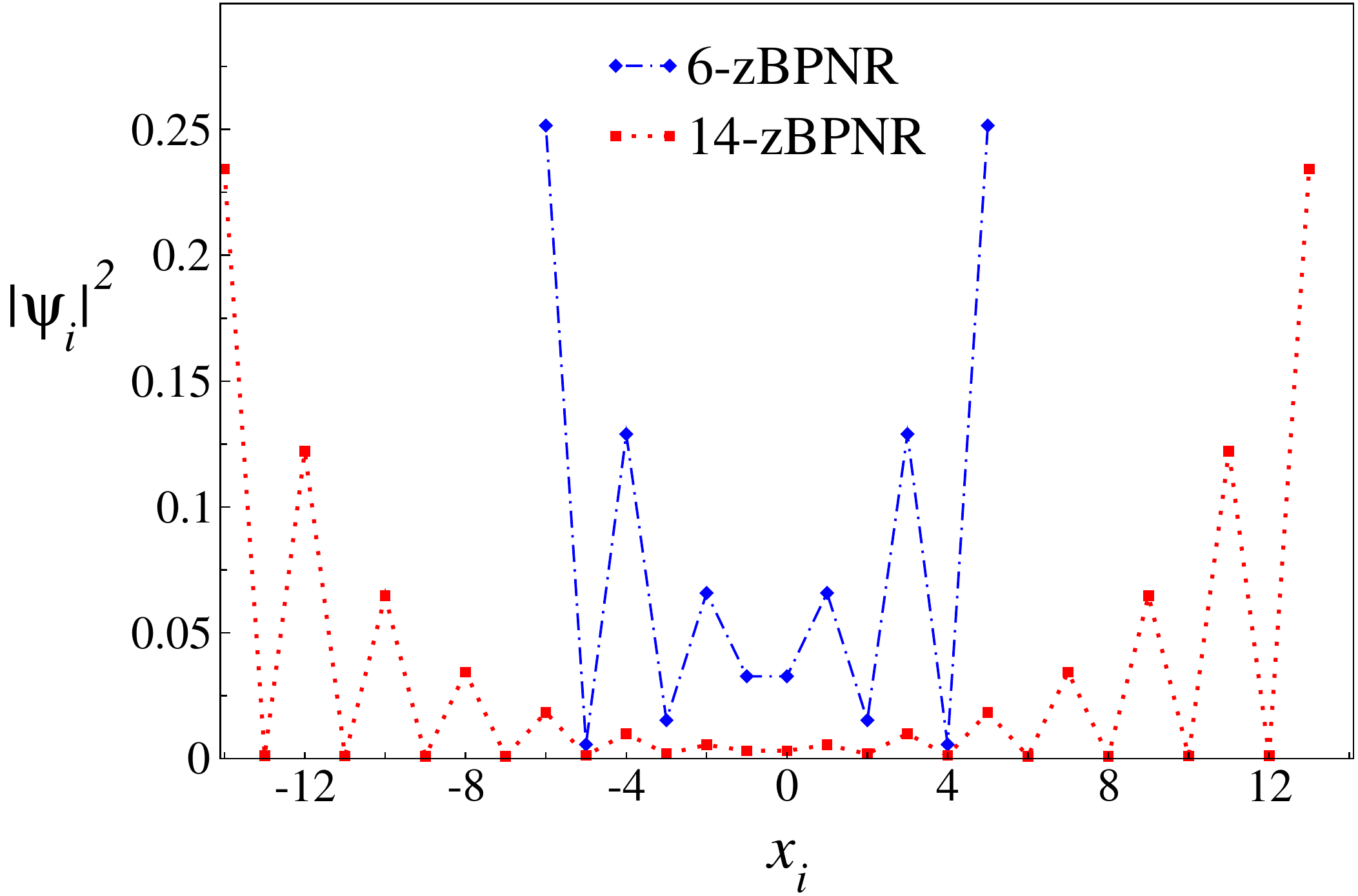}
\caption{Probability amplitude of the edge band eigenstates of 6-zPNR ($w\sim$ 1.25~nm) and 14-zPNR ($w\sim$ 3~nm). Note that the horizontal axis indicates a unit cell in the width of the ribbon.}
\label{fig:edgestate}
\end{figure}

\subsection{\label{EdgemodeszPNRs}Edge modes in zPNRs}
In order to understand the physics of this model, we study the influence of the ratio of the two dominant hopping parameters on the behavior of the electronic structure for zPNRs. 
We first study the dependence of quasi-flat bands and their corresponding edge states in zPNRs on the ratio of $|t_2/t_1|$. 
The band structure and probability amplitude of the upper valence band eigenstates of 100-zPNRs 
for $|t_2/t_1|=$1, 2, and 3 for $k=0$ are shown in Fig.~\ref{fig:t2t1}.
As can be seen in Figs.~\ref{fig:t2t1}(a), (b), and (c), as the $|t_2/t_1|$ ratio increases, the two middle bands (shown with grey lines) are detached from the bulk bands. The critical value of the ratio for the emergence of edge states at $k=0$ is 2, namely, at this ratio, the average amplitude of $|\Psi_{i}|^{2}$ becomes nearly homogeneous in the bulk. It should be noted that the states corresponding to the quasi-flat bands that are outside the middle region including between Dirac-like points and $k=\pi$ or $k=-\pi$ are always localized on the edges. Fig.~\ref{fig:t2t1}(c) shows the band structure for $|t_2/t_1|$=3. In this case, the edge bands are isolated from the bulk states, and are two-fold degenerate.
This degeneracy is lifted in zPNRs with small widths ($N_{z}<40$) for wave vectors near $k=0$.
This behavior can be explained by considering the effect of finite electron tunneling between two opposite edges of zPNRs with small widths~\cite{Dolui2012}.
Fig.~\ref{fig:t2t1}(d) shows the probability amplitude of the upper valence band eigenstate for $k=0$ as a function of the position of phosphorus atoms. As can be clearly seen, for small values of this ratio, the probability amplitude is large for the bulk sites, whereas for the edge sites, it is minimal or zero.
The probability amplitude of the bulk sites decreases as this ratio is increased. For $|t_2/t_1|$=3, only the probability amplitude of the sites near the edges are non-zero, thus indicating the importance of the $|t_2/t_1|$ ratio in the creation of the edge states. 

If we refer to the model that was introduced in Fig.~\ref{fig:eq_model}, we can explain the above mentioned behavior as follows. The structure shown in Fig.~\ref{fig:eq_model} is a bipartite lattice, and each site is connected to three sites of the other sublattice with two $t_1$ links and one $t_2$ link. If we separate the total wave function to two components, each having amplitudes only on one sublattice, the local energy contribution of a wave function is proportional to the local amplitudes of the two component wave functions times $\Delta\equiv 2t_1+t_2$. In the case of $\Delta<0$ ($|t_2/t_1|<2$), it is energetically more favourable for the two component wave functions to have maximum overlap, whereas in the case of $\Delta>0$ ($|t_2/t_1|>2$), we expect the two component waves to repel each other and push each other to the two edges of the nanoribbon. This is consistent with what is shown in Fig.~\ref{fig:t2t1}(d). It should be mentioned that the above discussion is only valid for the small values of the wave vector along the armchair direction, which corresponds to the quasi-flat bands at the Fermi level.
 
Clearly, the ribbon width is also important for the creation of the edge states in zPNRs. In zPNRs, the ribbon width must be greater than around 3 nm, which corresponds to 14-zPNR, for the edge states to appear. 
Fig.~\ref{fig:edgestate} shows the squared wave functions of the states in the edge band of 6-zPNR and 14-zPNR. 
For zPNRs with widths greater than 3 nm, the wave function corresponding to the two edges starts to decouple and will localize on the opposite edges.

\subsection{\label{Scaling_laws_PNRs}Scaling laws of band gaps for PNRs}
\begin{figure} 
\centering
\vspace{20pt}
\includegraphics[width=0.45\textwidth]{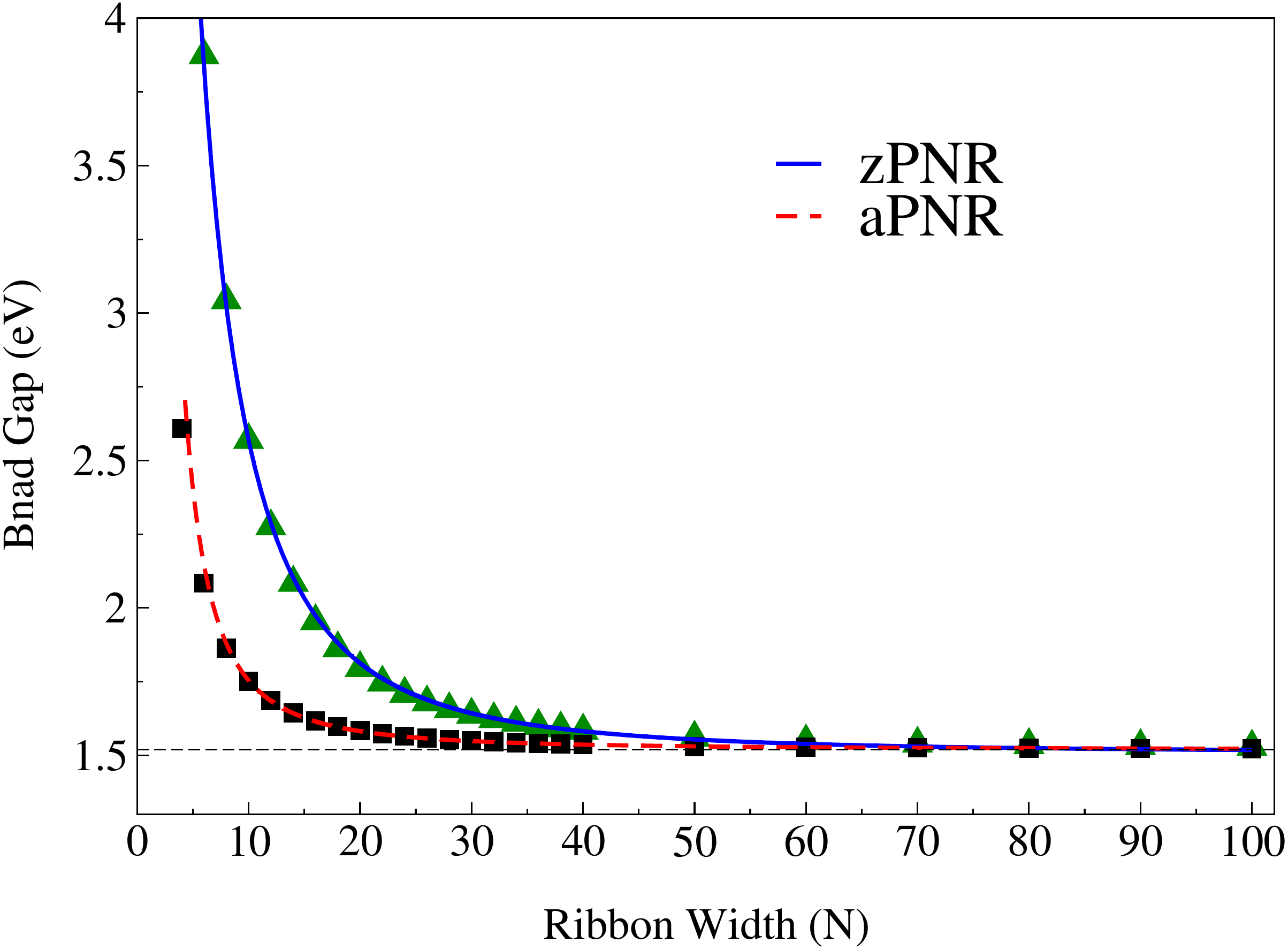}
\caption{Variation in band gap of zPNRs and aPNRs with ribbon width.}
\label{fig:bandgap}
\end{figure}

Fig.~\ref{fig:bandgap} shows the variation in band gap with ribbon width for zPNRs and aPNRs owing to the quantum confinement effect~\cite{Son20061,Zhao2004,Yang20071,Park2008}. 
In contrast to boron nitride nanoribbons (BNNRs)~\cite{Zhang2008}, graphene nanoribbons (GNRs)~\cite{Son20061}, and $\alpha$-graphdiyne nanoribbons~\cite{Niu2014}, the band gap of PNRs decreases monotonically as the ribbon width increases. 

\begin{figure} 
\centering
\vspace{20pt}
\includegraphics[width=0.45\textwidth]{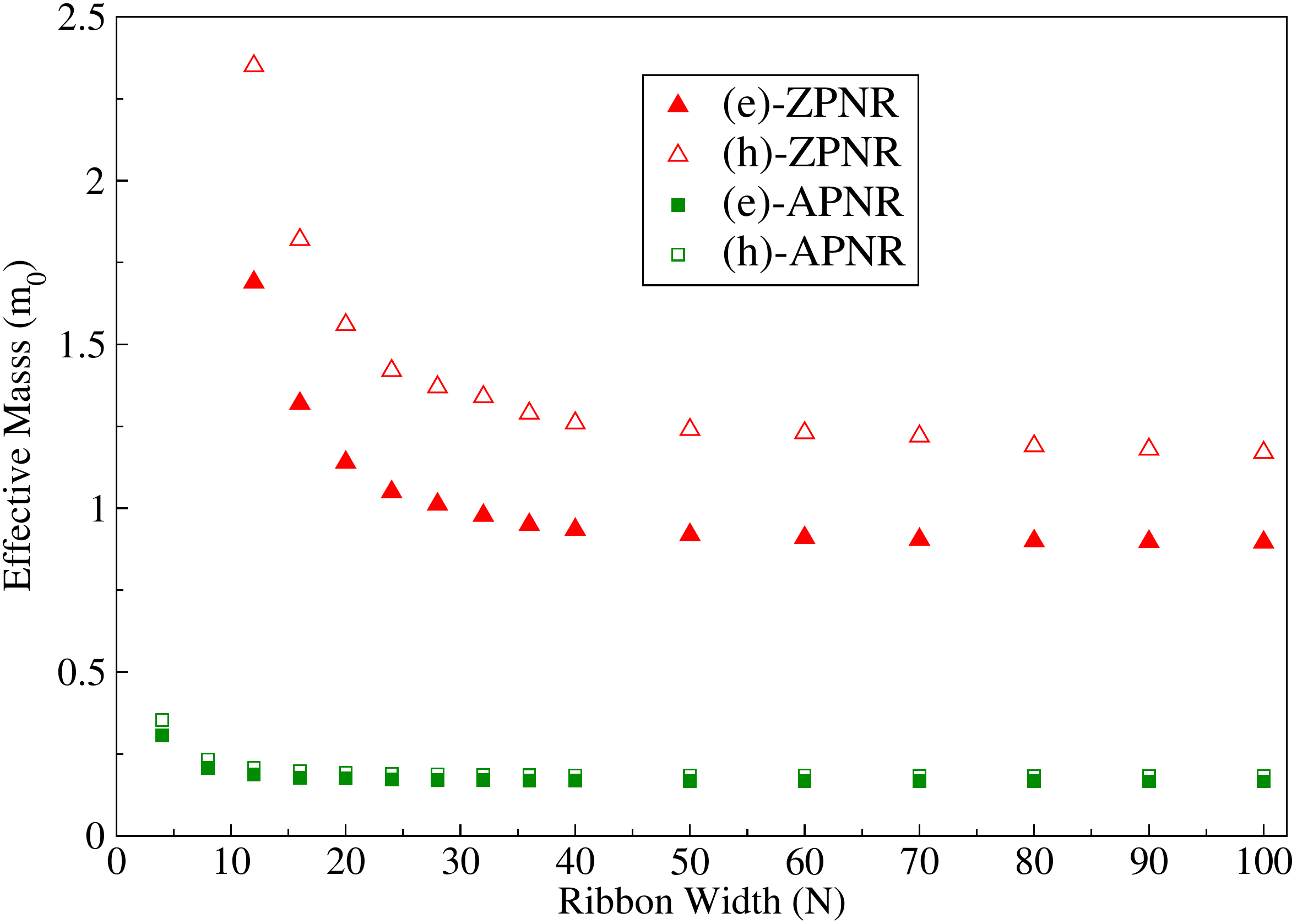}
\caption{Variation in effective masses of zPNRs and aPNRs with ribbon width.}
\label{fig:effmass}
\end{figure}

Fig.~\ref{fig:bandgap} shows that the bang gap is larger in zPNRs for the same ribbon width, indicating that the energy contribution from quantum confinement is higher in zPNRs, thus resulting in a stronger quantum confinement effect in zPNRs.  
The scaling behavior of band gap with increasing ribbon width for both types of PNRs has been calculated using DFT calculations~\cite{Tran2014,Wu2014}. They suggested a scaling behavior of $\sim 1/w^{2}$ for aPNRs whereas a $\sim 1/w$ for zPNRs. We argue that the scaling law for the zPNRs is not $1/w$. In fact, since the electrons along the confinement direction of zPNRs, which is the armchair direction, behave like massive-relativistic particles, we fit our data for zPNRs with $E_{gap} = \sqrt{A^2/{N_z}^{\alpha}+B^2}+C$ ($w\simeq 0.22N_z-0.08$~nm). The fitted values for the parameters are $A=22.9$~eV, $\alpha=2.18$, $B=1.10$~eV, and $C=0.42$~eV. In this formula, we expect a parabolic scaling law as long as the second term under the square root is much larger than the first term. This condition for the above fitted values occurs for $w\gg 3.5$~nm. In the massless-relativistic limit, where the energy contribution to the quantum confinement is proportional to the momentum, the first term is much larger than the second term, and this condition occurs for $w\ll 3.5$~nm. It should be mentioned that the band dispersion of the DFT calculations near the gap, specially for the $\Gamma$-X direction, are very close to our TB calculations, and the above discussion is also applicable to their scaling graphs. The maximum widths considered in the DFT calculations for the scaling is 3~nm; therefore, they have not been able to consider the parabolic region. According to the above discussion, we should not expect a $1/w$ scaling law for zPNRs with ribbon widths larger than $3.5$~nm. 
For aPNRs we fit the data with $E_{gap} = A'/{N_a}^\beta + C'$ ($w\simeq 0.164(N_a-1)$~nm), and the fitted values for the parameters are $A'=20.4$~eV, $\beta=1.92$, and $C'=1.52$~eV, in agreement with previous results~\cite{Tran2014,Wu2014}.

We have also calculated the effective masses of the electron and hole states near the VBM and CBM of PNRs with different ribbon widths. The results are shown if Fig.~\ref{fig:effmass}. The effective masses of zPNRs are more than six times larger than aPNRs and for small widths their effective masses increase even to higher values. 

\begin{figure}[b]
\vspace{20pt}
\includegraphics[width=0.45\textwidth]{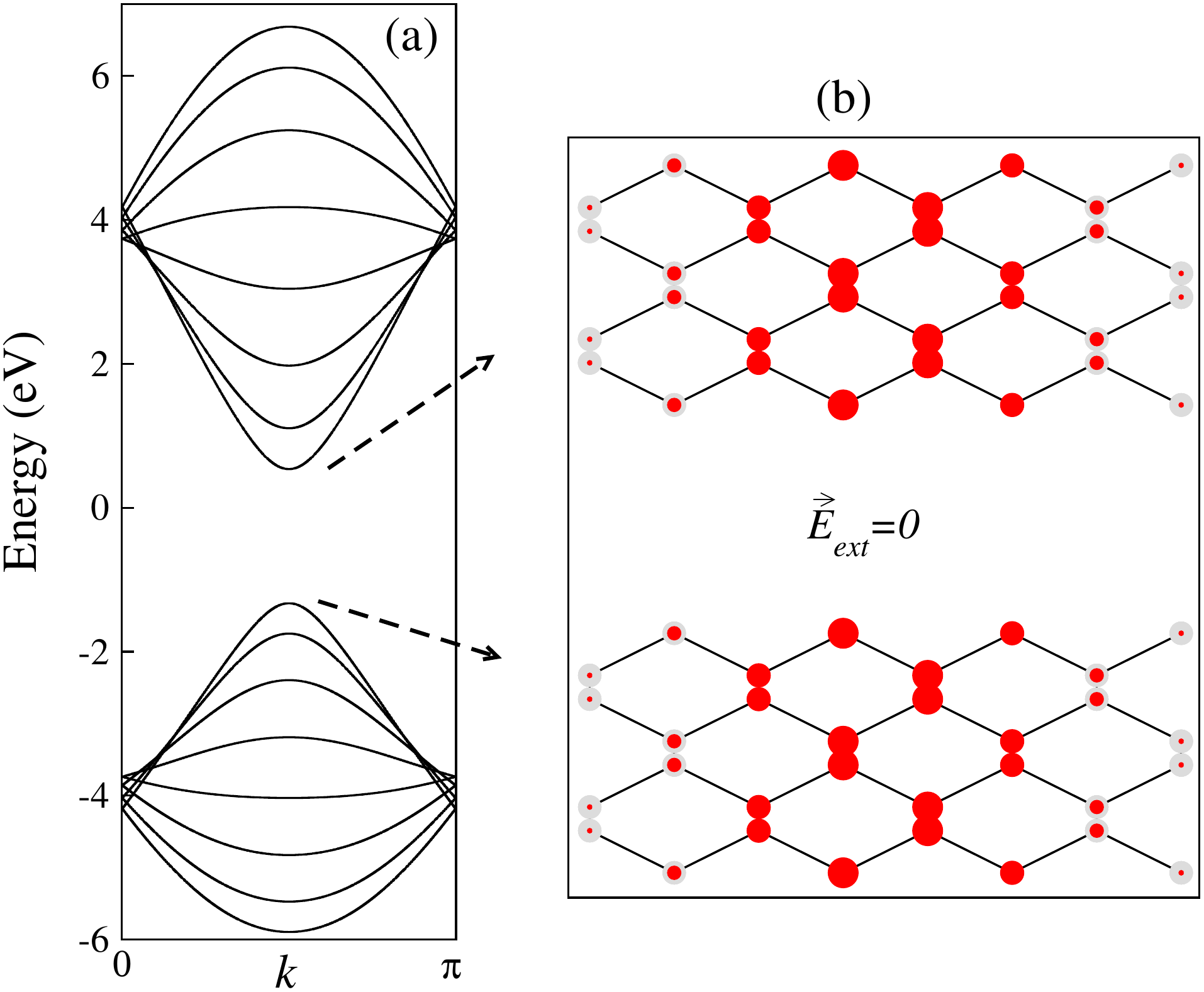}

\vspace{35pt}
\includegraphics[width=0.45\textwidth]{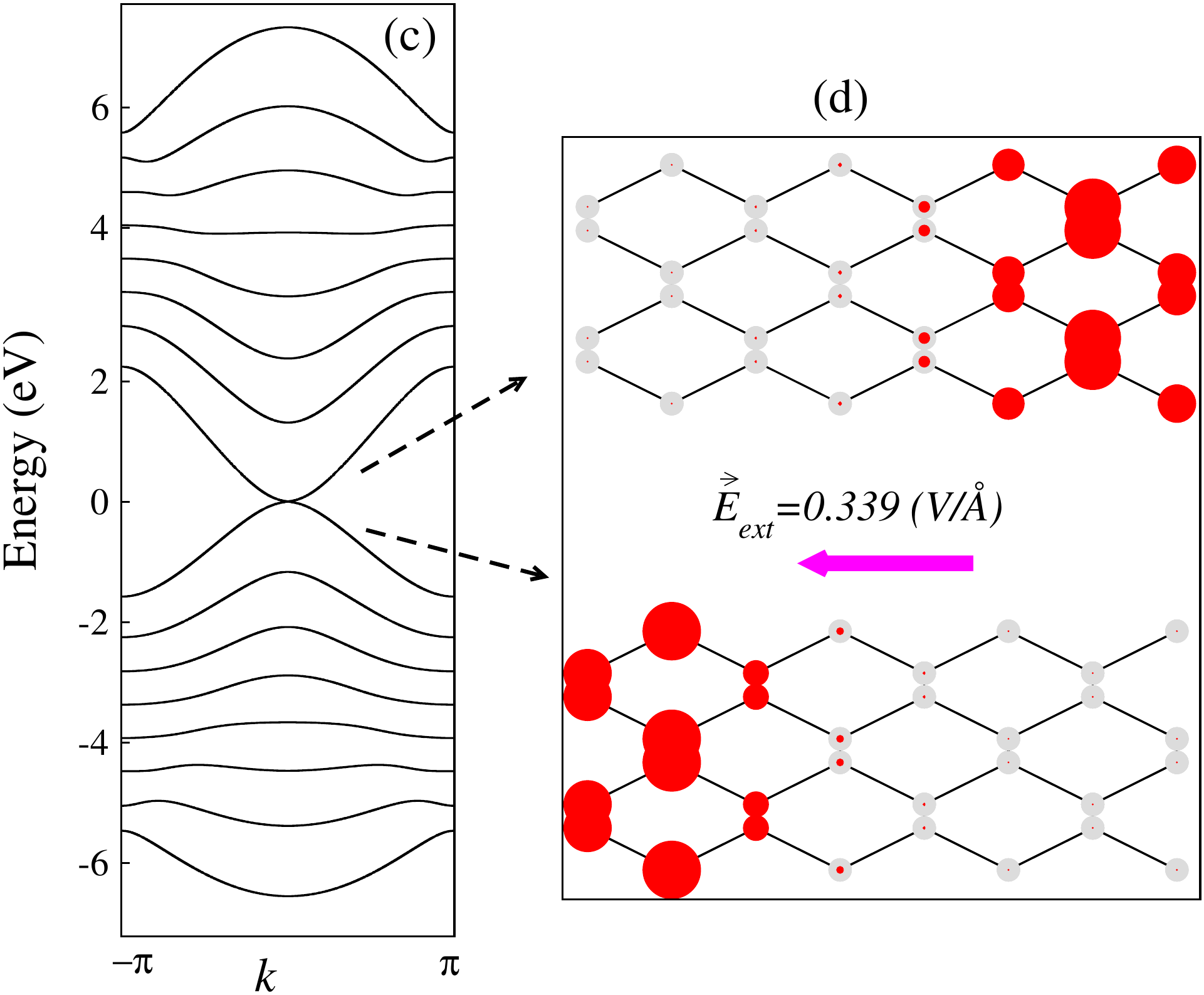}
\caption{Top: (a) Band structure, and (b) probability amplitudes of 8-aPNR for zero transverse electric field. Bottom: The same graph for $E_{ext}=0.339$~V/\AA. Note that the eigenstates correspond to $k=0$.}
 \label{fig:armEband}
\end{figure}

%
%

\begin{figure} 
\centering
\vspace{15pt}
\includegraphics[width=0.45\textwidth]{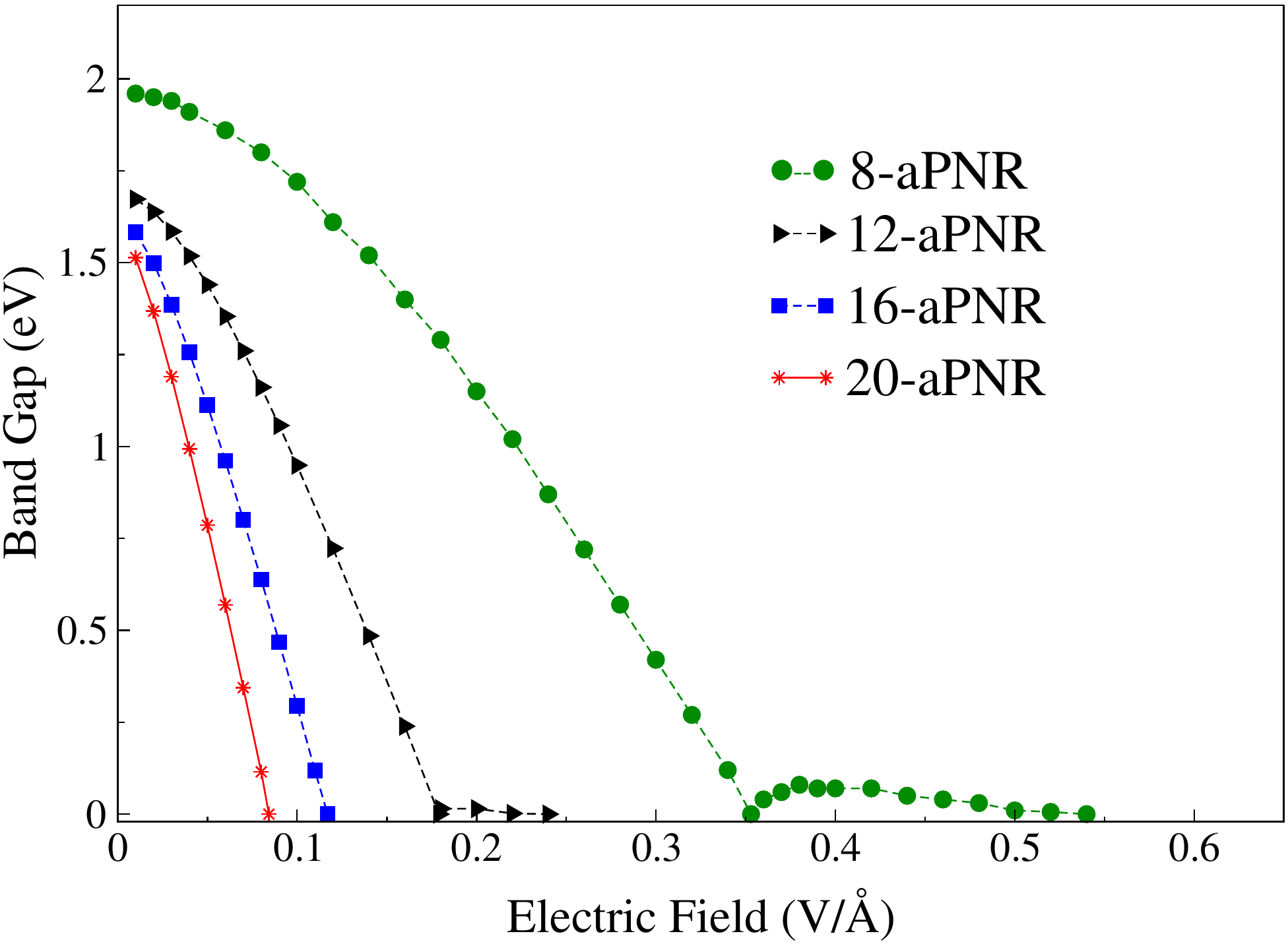}
\caption{Variation in band gap of aPNRs with transverse external electric field for five different ribbon widths.}
\label{fig:bandgapEy}
\end{figure}

\begin{figure} 
\centering
\vspace{20pt}
\includegraphics[width=0.35\textwidth]{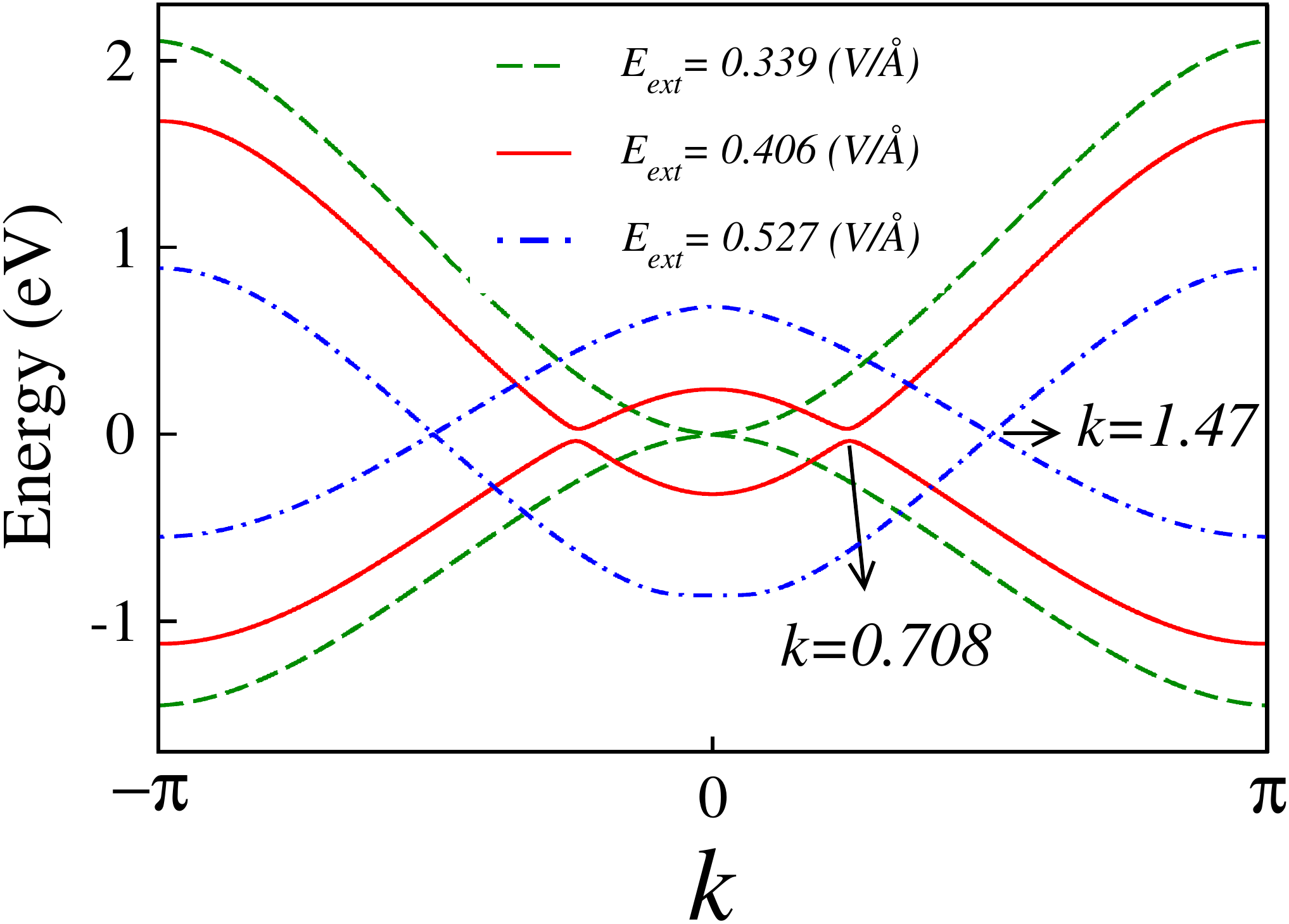}
\caption{Conduction and valance bands of 8-aPNR for $E_{ext}=0.339$, $0.406$, and $0.527$~V/\AA.}
\label{fig:psi8Eyarm}
\end{figure} 


\subsection{\label{Response_aPNRs}Response of aPNRs to ${\bf E_{ext}}$}
Next, we analyze the relationship between the electronic properties of aPNRs (periodicity along the $x$-direction) and the external electric field ($E_{ext}$) along the ribbon width. The band structure for $E_{ext}=0$ is shown in Fig.~\ref{fig:armEband}(a), in which the CBM and VBM determine the band gap. The electronic states associated with the VBM and CBM are located in the bulk of the ribbon [Fig.~\ref{fig:armEband}(b)]. Also, all aPNRs are semiconductors independent of their ribbon width. When a transverse $E_{ext}$ is applied along the width, the states corresponding to the CBM, which have a positive band curvature (electron states), will shift to lower energies owing to Stark effect, whereas the states corresponding to the VBM (hole states) shift to higher energies. Therefore, the CBM and VBM states will localize on the ribbon edges [Fig.~\ref{fig:armEband}(d)]. By further increasing the field strength, the two bands approach one another 
because of the electrostatic potential difference between the opposite edges, and the band gap decreases and eventually closes at a critical transverse field, $E_{c}$ [Fig.~\ref{fig:armEband}(c)]. 
This trend in band gap variation with $E_{ext}$ has already been observed in other materials such as GNRs~\cite{Son20061}, carbon nanotubes~\cite{Okeeffe2002}, MoS$_2$ nanoribbons~\cite{Dolui2012}, and BNNRs~\cite{Zhang2008,Park2008,Barone2008}.
It should be noted that in contrast to other compounds such as BN~\cite{Zhang2008},
that the structure have a polarization along the width, the gap closure does not change if we reverse the direction of the transverse $E_{ext}$ along the width.

We also calculated the variation in band gap of aPNRs with $E_{ext}$ for four different widths (Fig.~\ref{fig:bandgapEy}).
As the transverse $E_{ext}$ increases, the band gap decreases uniformly. Similar behavior has been observed in the nanoribbons of BN~\cite{Zhang2008,Park2008} and MoS$_2$~\cite{Dolui2012}.

As the aPNR width increases, the band gap decreases rapidly with increasing transverse field $E_c$, and the gap closure occurs for smaller fields because the electrostatic potential difference is proportional to the ribbon width. 
The variation in band gap with ribbon width and transverse $E_{ext}$ has been calculated recently using DFT~\cite{Wu2014}.   
For aPNRs with large widths, the results obtained with the TB approach are in good agreement with the DFT-calculations.
As the transverse $E_{ext}$ increases, the gap closes directly at $k=0$ for $E_{c}$=0.339 V/\AA, and the edge band states corresponding to the VBM and CBM states are localized on the opposite edges of the aPNRs [Fig.~\ref{fig:armEband}(d)].

As shown in Fig.~\ref{fig:bandgapEy}, the gap closure of aPNRs with small widths exhibits an interesting trend.
For instance, for the 8-aPNR, the band gap varies slowly under a strong $E_{ext}$, and the band gap closes for $E_{c}$=0.339, it opens again and closes at 0.527 V/\AA.
Fig.~\ref{fig:psi8Eyarm} shows the valence and conduction bands for $E_{ext}$= 0.339, 0.406, and 0.527 V/\AA.
The opening up of the band gap after its closure for very small ribbon widths is related to the finite hopping integrals between the two opposite edges and the mechanism for a similar behavior in MoS$_2$ nanoribbons has been explained elsewhere~\cite{Dolui2012}.

\begin{figure}[t]
\vspace{20pt}
\includegraphics[width=0.46\textwidth]{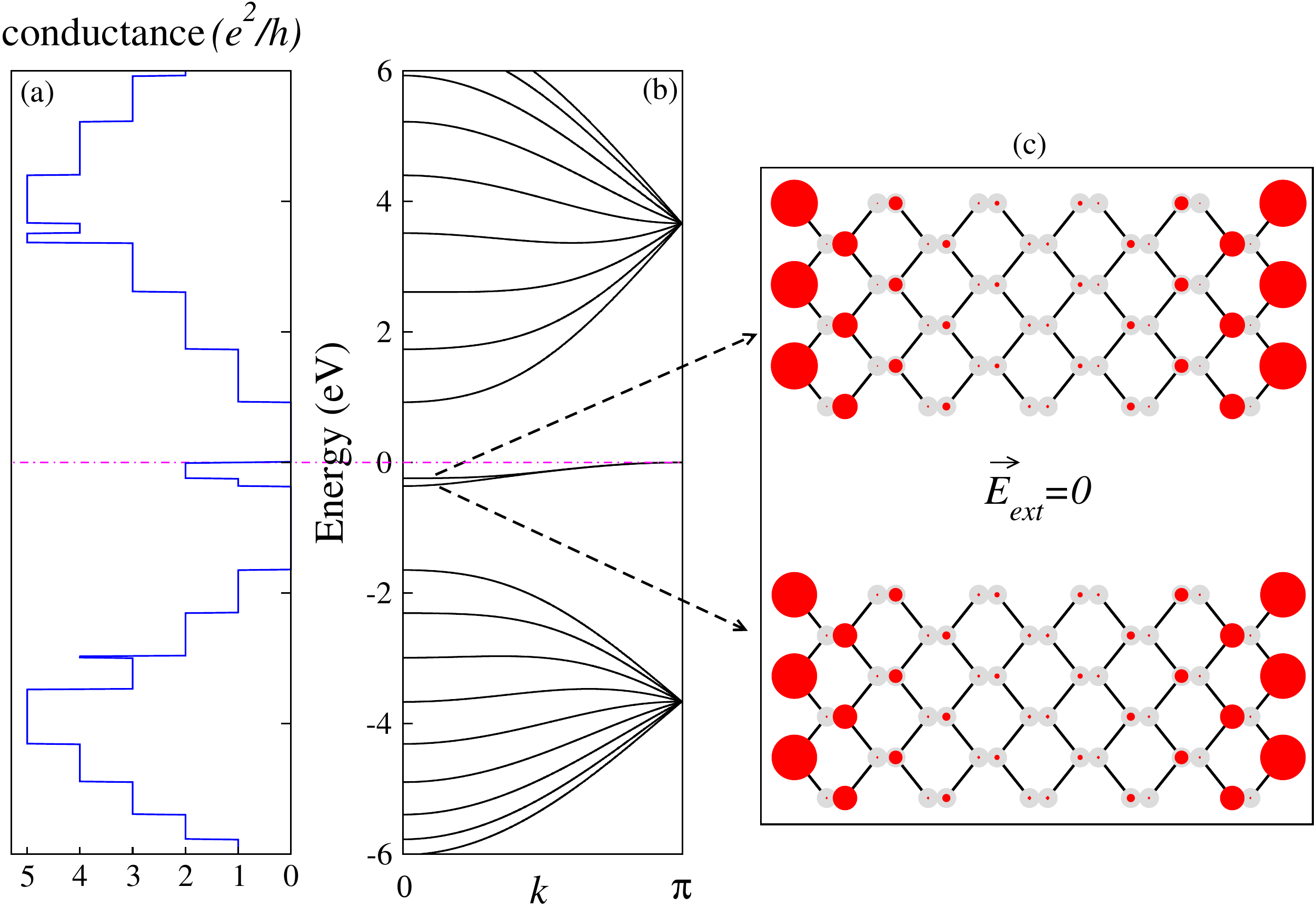}

\vspace{20pt}
\includegraphics[width=0.46\textwidth]{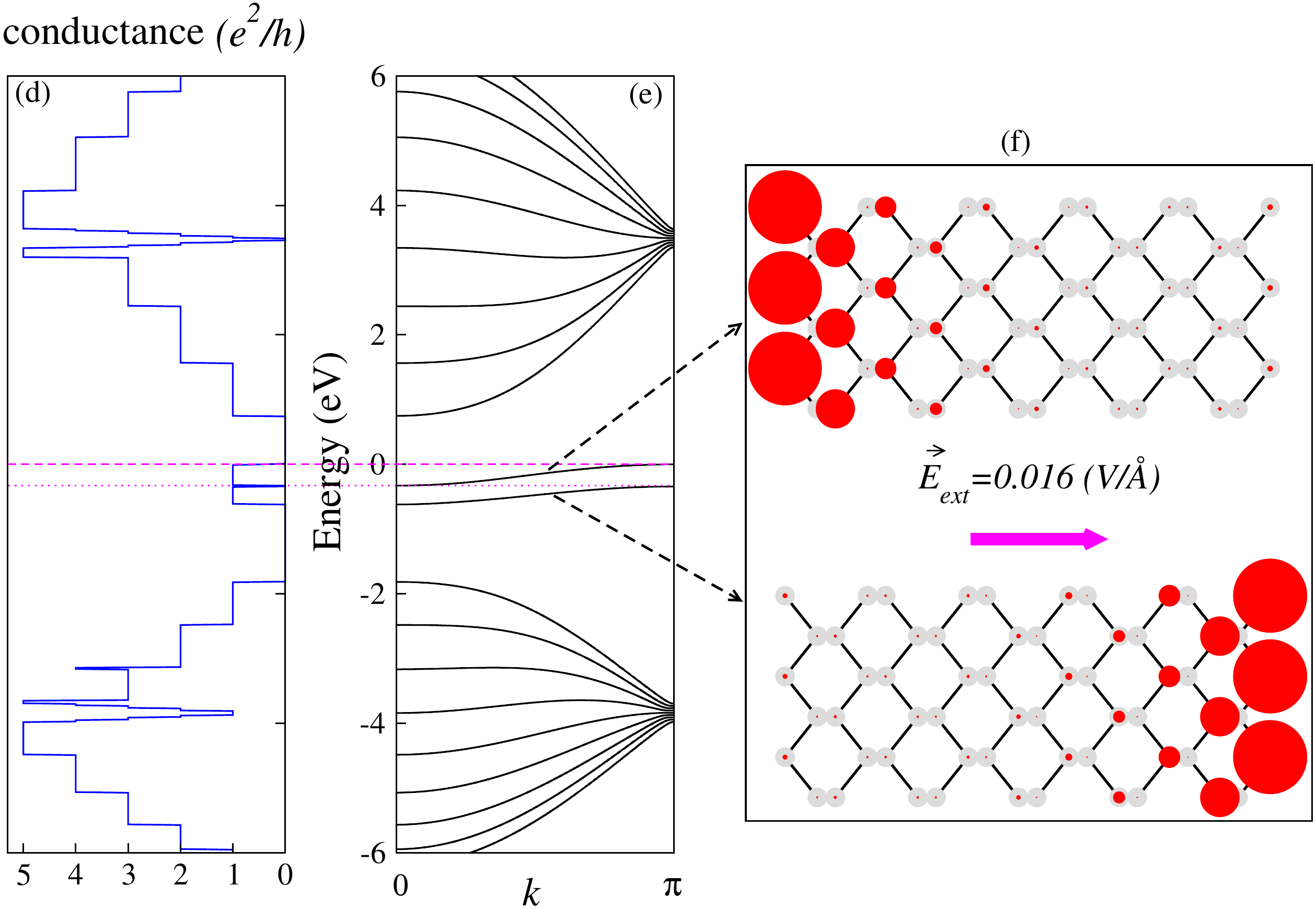}
\caption{Top: (a) Conductance, (b) band structure, and (c) probability amplitudes of the band gap edge states of a 10-zPNR under zero transverse electric field. Bottom: The same graph for $E_{ext}=0.016$~V/\AA. Note that the eigenstates correspond to $k=0$.}
 \label{fig:zigEband}
\end{figure}

\begin{figure}[t] 
\centering
\vspace{30pt}
\includegraphics[width=0.45\textwidth]{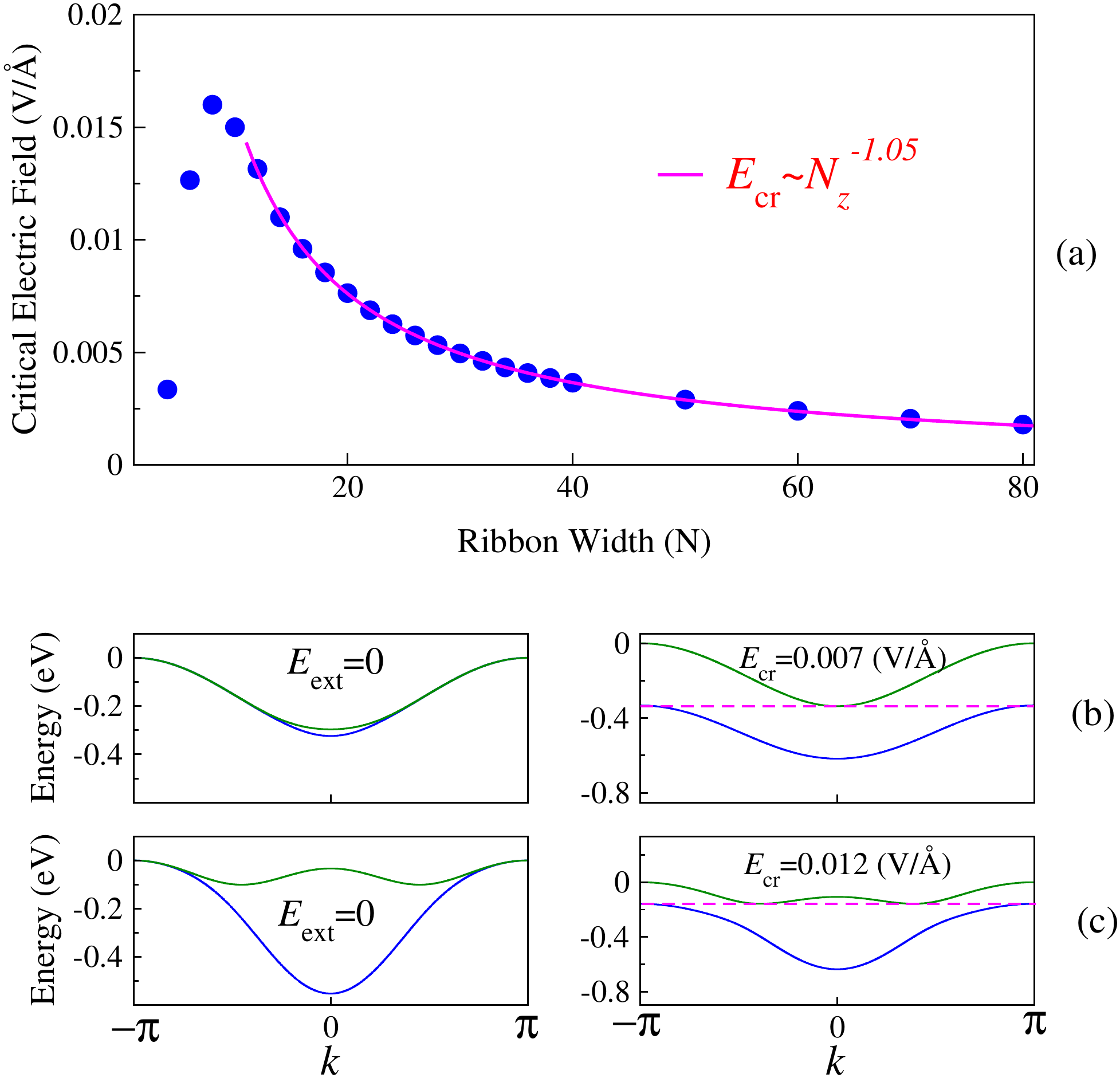}
\caption{(a) Variation in critical transverse electric field with ribbon width of zPNRs. Quasi-flat bands for (b) 20-zPNR for $E_{ext}=0$ and $E_{ext}=0.007$~V/\AA and (c) 6-zPNR for $E_{ext}=0$ and $E_{ext}=0.012$~V/\AA.}
\label{fig:zigEcN}
\end{figure} 

\subsection{\label{Transistor_zPNRs}Transistor effect in zPNRs}
A recent study based on the TB model has investigated the effect of an external in-plane ($E_{ext}$) electric field on the edge modes of zPNRs and the effect of an external electric field ($E_{z}$) perpendicular to the ribbon surface on zPNRs~\cite{Ezawa2014}. The results show that the band gap increases in accordance with ${(lE_{z})}^{2}$ where $l$ is the  separation distance between the upper and lower layers of phosphorene.
Moreover, for $E_{ext}$ greater than a critical strength $(E_c)$, the degeneracy of the edge bands in Fig.~\ref{fig:t2t1}c is lifted for the quasi-flat bands, and a transistor effect can be observed. Further, $E_c$ is inversely proportional to the ribbon width ($\propto 1/w$).

In this study, we investigated the transistor effect in zPNRs using the Landauer formalism~\cite{Datta1995,Datta2005}.
In this formalism, the conductance ${\sigma(E)}$ for nanoscale devices at Fermi energy ($E_F$) between a pair of leads $p$ and $q$ is given by
\begin{equation}
 \sigma(E)=(\frac{e^2}{h})Tr[\Gamma_{p}(E)G^{R}_{D}(E)\Gamma_{q}(E)G^{A}_{D}(E)]
\end{equation}
where $G^{R}_{D}(E)$ is the retarded Green's function of the device and $G^{A}_{D}(E)={G^{R}_{D}}^{\dagger}(E)$. 
In this equation, $\Gamma_{p(q)}=i[\Sigma_{p(q)}(E)-{\Sigma_{p(q)}}^{\dagger}(E)]$ where $\Sigma_{p(q)}(E)$ is the self energy related to lead $p$ ($q$).  The retarded Green's function of the device ($G^{R}_{D}(E)$) is given by
\begin{equation}
 G^{R}_{D}(E)=[E-H_{D}-\Sigma^{R}_p(E)-\Sigma^{R}_q(E)]^{-1}
\end{equation}

We now analyze the conditions under which the transistor effect can be observed in zPNRs.
The conductance, band structure, and wave functions of a 10-zPNRs for $E_{ext}$=0 and 0.016 V/\AA~ are shown in Fig.~\ref{fig:zigEband}.
As can be seen in Fig.~\ref{fig:zigEband}(b), the degeneracy between the two edge modes at zPNRs is slightly lifted close to $k=0$. Therefore, the conductance is slightly asymmetric near $k=0$.
As shown in Fig.~\ref{fig:zigEband}(c), the wave functions of the upper and lower quasi-flat bands are localized on both the edges.
As the external electric field is increased up to a critical field, the overlap between these two bands vanishes. What we have is a conductance controlled by the external electric field at Fermi energy, which is a field-effect transistor behavior.  
In this case, the wave functions of the upper and lower edge bands are localized on the opposite edges [Fig.~\ref{fig:zigEband}(f)]. 

The relationship between $E_{c}$ and ribbon width is shown in Fig.~\ref{fig:zigEcN}(a).
For zPNRs with widths greater than $N_{z}=14$, $E_{c}$ scales as ${1/{N_z}^{1.05}}$, which is in good agreement with the results previously reported by Ezawa~\cite{Ezawa2014}. However, for ribbons with widths smaller than $N_{z}=14$, we found a completely different behavior. To explain this portion of the graph, we considered the behavior of the edge bands of zPNRs with different widths.
Figs.~\ref{fig:zigEcN}(b) and \ref{fig:zigEcN}(c) show the quasi-flat bands for 20-zPNR and 6-zPNR, respectively.
For $E_{ext}=0$, the quasi-flat bands are different for these two widths. The VBM and CBM of 20-zPNR are located at $k=\pi$ and $k=0$, respectively.
The VBM of 6-zPNR is also located at $k=\pi$ whereas the CBM is located at a $k$ between 0 and $\pi$.
This displacement of the CBM in 6-zPNR is caused by the finite interaction between the two edge modes.
Therefore, a lower external electric field is needed for observing the transistor effect.  

\section{\label{Conclusion}Conclusion}
In summary, we presented the numerical results for the band structure and quantum conductance of zPNRs and aPNRs based on a five parameter TB model. It was shown that the general form of the electronic structure is controlled by the two dominant hopping parameters. It was discussed that the opposite sign of these two hopping integrals is the origin of the creation of a relativistic band dispersion along the armchair direction. Our numerical results for zPNRs predicts a pair of degenerate quasi-flat bands at the Fermi level that are localized on the ribbon edges, and this degeneracy is lifted for small ribbon widths owing to finite interactions between the edge states. Additionally, our calculations provide scaling laws of the band gap for PNRs as a function of ribbon width. We discussed that the band gap scaling law for both nanoribbons with widths much larger than $3.5$~nm is always $1/w^2$. For aPNRs, a semiconducting behavior is predicted, and an insulator-metal transition can be expected when a transverse electric field is applied. In zPNRs, an external transverse electric field can remove the overlap between quasi-flat bands. The anisotropy in the mobility , tunability of the band gap with ribbon width, and the field dependent conductance make this system a promising candidate for the future of field-effect transistor technologies.

\nocite{*}

\bibliography{our_TZF_article1}

\begin{thebibliography}{44}%
\makeatletter
\providecommand \@ifxundefined [1]{%
 \@ifx{#1\undefined}
}%
\providecommand \@ifnum [1]{%
 \ifnum #1\expandafter \@firstoftwo
 \else \expandafter \@secondoftwo
 \fi
}%
\providecommand \@ifx [1]{%
 \ifx #1\expandafter \@firstoftwo
 \else \expandafter \@secondoftwo
 \fi
}%
\providecommand \natexlab [1]{#1}%
\providecommand \enquote  [1]{``#1''}%
\providecommand \bibnamefont  [1]{#1}%
\providecommand \bibfnamefont [1]{#1}%
\providecommand \citenamefont [1]{#1}%
\providecommand \href@noop [0]{\@secondoftwo}%
\providecommand \href [0]{\begingroup \@sanitize@url \@href}%
\providecommand \@href[1]{\@@startlink{#1}\@@href}%
\providecommand \@@href[1]{\endgroup#1\@@endlink}%
\providecommand \@sanitize@url [0]{\catcode `\\12\catcode `\$12\catcode
  `\&12\catcode `\#12\catcode `\^12\catcode `\_12\catcode `\%12\relax}%
\providecommand \@@startlink[1]{}%
\providecommand \@@endlink[0]{}%
\providecommand \url  [0]{\begingroup\@sanitize@url \@url }%
\providecommand \@url [1]{\endgroup\@href {#1}{\urlprefix }}%
\providecommand \urlprefix  [0]{URL }%
\providecommand \Eprint [0]{\href }%
\providecommand \doibase [0]{http://dx.doi.org/}%
\providecommand \selectlanguage [0]{\@gobble}%
\providecommand \bibinfo  [0]{\@secondoftwo}%
\providecommand \bibfield  [0]{\@secondoftwo}%
\providecommand \translation [1]{[#1]}%
\providecommand \BibitemOpen [0]{}%
\providecommand \bibitemStop [0]{}%
\providecommand \bibitemNoStop [0]{.\EOS\space}%
\providecommand \EOS [0]{\spacefactor3000\relax}%
\providecommand \BibitemShut  [1]{\csname bibitem#1\endcsname}%
\let\auto@bib@innerbib\@empty
\bibitem [{\citenamefont {Novoselov}\ \emph {et~al.}(2004)\citenamefont
  {Novoselov}, \citenamefont {Geim}, \citenamefont {Morozov}, \citenamefont
  {Jiang}, \citenamefont {Zhang}, \citenamefont {Dubonos}, \citenamefont
  {Grigorieva},\ and\ \citenamefont {Firsov}}]{Novoselov2004}%
  \BibitemOpen
  \bibfield  {author} {\bibinfo {author} {\bibfnamefont {K.~S.}\ \bibnamefont
  {Novoselov}}, \bibinfo {author} {\bibfnamefont {A.~K.}\ \bibnamefont {Geim}},
  \bibinfo {author} {\bibfnamefont {S.~V.}\ \bibnamefont {Morozov}}, \bibinfo
  {author} {\bibfnamefont {D.}~\bibnamefont {Jiang}}, \bibinfo {author}
  {\bibfnamefont {Y.}~\bibnamefont {Zhang}}, \bibinfo {author} {\bibfnamefont
  {S.~V.}\ \bibnamefont {Dubonos}}, \bibinfo {author} {\bibfnamefont {I.~V.}\
  \bibnamefont {Grigorieva}}, \ and\ \bibinfo {author} {\bibfnamefont {A.~A.}\
  \bibnamefont {Firsov}},\ }\href@noop {} {\bibfield  {journal} {\bibinfo
  {journal} {Science}\ }\textbf {\bibinfo {volume} {306}},\ \bibinfo {pages}
  {666} (\bibinfo {year} {2004})}\BibitemShut {NoStop}%
\bibitem [{\citenamefont {Geim}\ and\ \citenamefont
  {Novoselov}(2007)}]{Geim2007}%
  \BibitemOpen
  \bibfield  {author} {\bibinfo {author} {\bibfnamefont {A.~K.}\ \bibnamefont
  {Geim}}\ and\ \bibinfo {author} {\bibfnamefont {K.~S.}\ \bibnamefont
  {Novoselov}},\ }\href@noop {} {\bibfield  {journal} {\bibinfo  {journal}
  {Nature Materials}\ }\textbf {\bibinfo {volume} {6}},\ \bibinfo {pages} {183}
  (\bibinfo {year} {2007})}\BibitemShut {NoStop}%
\bibitem [{\citenamefont {Neto}\ \emph {et~al.}(2009)\citenamefont {Neto},
  \citenamefont {Guinea}, \citenamefont {Peres}, \citenamefont {Novoselov},\
  and\ \citenamefont {Geim}}]{Neto2009}%
  \BibitemOpen
  \bibfield  {author} {\bibinfo {author} {\bibfnamefont {A.~H.~C.}\
  \bibnamefont {Neto}}, \bibinfo {author} {\bibfnamefont {F.}~\bibnamefont
  {Guinea}}, \bibinfo {author} {\bibfnamefont {N.~M.~R.}\ \bibnamefont
  {Peres}}, \bibinfo {author} {\bibfnamefont {K.~S.}\ \bibnamefont
  {Novoselov}}, \ and\ \bibinfo {author} {\bibfnamefont {A.~K.}\ \bibnamefont
  {Geim}},\ }\href@noop {} {\bibfield  {journal} {\bibinfo  {journal} {Rev.
  Mod. Phys.}\ }\textbf {\bibinfo {volume} {81}},\ \bibinfo {pages} {109}
  (\bibinfo {year} {2009})}\BibitemShut {NoStop}%
\bibitem [{\citenamefont {Splendiani}\ \emph {et~al.}(2010)\citenamefont
  {Splendiani}, \citenamefont {Sun}, \citenamefont {Zhang}, \citenamefont {Li},
  \citenamefont {Kim}, \citenamefont {Chim}, \citenamefont {Galli},\ and\
  \citenamefont {Wang}}]{Splendiani2010}%
  \BibitemOpen
  \bibfield  {author} {\bibinfo {author} {\bibfnamefont {A.}~\bibnamefont
  {Splendiani}}, \bibinfo {author} {\bibfnamefont {L.}~\bibnamefont {Sun}},
  \bibinfo {author} {\bibfnamefont {Y.}~\bibnamefont {Zhang}}, \bibinfo
  {author} {\bibfnamefont {T.}~\bibnamefont {Li}}, \bibinfo {author}
  {\bibfnamefont {J.}~\bibnamefont {Kim}}, \bibinfo {author} {\bibfnamefont
  {C.~Y.}\ \bibnamefont {Chim}}, \bibinfo {author} {\bibfnamefont
  {G.}~\bibnamefont {Galli}}, \ and\ \bibinfo {author} {\bibfnamefont
  {F.}~\bibnamefont {Wang}},\ }\href@noop {} {\bibfield  {journal} {\bibinfo
  {journal} {Nano Lett.}\ }\textbf {\bibinfo {volume} {10}},\ \bibinfo {pages}
  {1271} (\bibinfo {year} {2010})}\BibitemShut {NoStop}%
\bibitem [{\citenamefont {Mak}\ \emph {et~al.}(2010)\citenamefont {Mak},
  \citenamefont {Lee}, \citenamefont {Hone}, \citenamefont {Shan},\ and\
  \citenamefont {Heinz}}]{Mak2010}%
  \BibitemOpen
  \bibfield  {author} {\bibinfo {author} {\bibfnamefont {K.~F.}\ \bibnamefont
  {Mak}}, \bibinfo {author} {\bibfnamefont {C.}~\bibnamefont {Lee}}, \bibinfo
  {author} {\bibfnamefont {J.}~\bibnamefont {Hone}}, \bibinfo {author}
  {\bibfnamefont {J.}~\bibnamefont {Shan}}, \ and\ \bibinfo {author}
  {\bibfnamefont {T.~F.}\ \bibnamefont {Heinz}},\ }\href@noop {} {\bibfield
  {journal} {\bibinfo  {journal} {Phys. Rev. Lett.}\ }\textbf {\bibinfo
  {volume} {105}},\ \bibinfo {pages} {136805} (\bibinfo {year}
  {2010})}\BibitemShut {NoStop}%
\bibitem [{\citenamefont {Xiao}\ \emph {et~al.}(2012)\citenamefont {Xiao},
  \citenamefont {Liu}, \citenamefont {Feng}, \citenamefont {Xu},\ and\
  \citenamefont {Yao}}]{Xiao2010}%
  \BibitemOpen
  \bibfield  {author} {\bibinfo {author} {\bibfnamefont {D.}~\bibnamefont
  {Xiao}}, \bibinfo {author} {\bibfnamefont {G.~B.}\ \bibnamefont {Liu}},
  \bibinfo {author} {\bibfnamefont {W.}~\bibnamefont {Feng}}, \bibinfo {author}
  {\bibfnamefont {X.}~\bibnamefont {Xu}}, \ and\ \bibinfo {author}
  {\bibfnamefont {W.}~\bibnamefont {Yao}},\ }\href@noop {} {\bibfield
  {journal} {\bibinfo  {journal} {Phys. Rev. Lett.}\ }\textbf {\bibinfo
  {volume} {108}},\ \bibinfo {pages} {196802} (\bibinfo {year}
  {2012})}\BibitemShut {NoStop}%
\bibitem [{\citenamefont {Blase}\ \emph {et~al.}(1995)\citenamefont {Blase},
  \citenamefont {Rubio}, \citenamefont {Louie},\ and\ \citenamefont
  {Cohen}}]{Blase1995}%
  \BibitemOpen
  \bibfield  {author} {\bibinfo {author} {\bibfnamefont {X.}~\bibnamefont
  {Blase}}, \bibinfo {author} {\bibfnamefont {A.}~\bibnamefont {Rubio}},
  \bibinfo {author} {\bibfnamefont {S.~G.}\ \bibnamefont {Louie}}, \ and\
  \bibinfo {author} {\bibfnamefont {M.~L.}\ \bibnamefont {Cohen}},\ }\href@noop
  {} {\bibfield  {journal} {\bibinfo  {journal} {Phys. Rev. B}\ }\textbf
  {\bibinfo {volume} {51}},\ \bibinfo {pages} {6868} (\bibinfo {year}
  {1995})}\BibitemShut {NoStop}%
\bibitem [{\citenamefont {Watanabe}\ \emph {et~al.}(2004)\citenamefont
  {Watanabe}, \citenamefont {Taniguchi},\ and\ \citenamefont
  {Kanda}}]{Watanabe2004}%
  \BibitemOpen
  \bibfield  {author} {\bibinfo {author} {\bibfnamefont {K.}~\bibnamefont
  {Watanabe}}, \bibinfo {author} {\bibfnamefont {T.}~\bibnamefont {Taniguchi}},
  \ and\ \bibinfo {author} {\bibfnamefont {H.}~\bibnamefont {Kanda}},\
  }\href@noop {} {\bibfield  {journal} {\bibinfo  {journal} {Nature Materials}\
  }\textbf {\bibinfo {volume} {3}},\ \bibinfo {pages} {404} (\bibinfo {year}
  {2004})}\BibitemShut {NoStop}%
\bibitem [{\citenamefont {Kuc}\ \emph {et~al.}(2011)\citenamefont {Kuc},
  \citenamefont {Zibouche},\ and\ \citenamefont {Heine}}]{Kuc2011}%
  \BibitemOpen
  \bibfield  {author} {\bibinfo {author} {\bibfnamefont {A.}~\bibnamefont
  {Kuc}}, \bibinfo {author} {\bibfnamefont {N.}~\bibnamefont {Zibouche}}, \
  and\ \bibinfo {author} {\bibfnamefont {T.}~\bibnamefont {Heine}},\
  }\href@noop {} {\bibfield  {journal} {\bibinfo  {journal} {Phys. Rev. B}\
  }\textbf {\bibinfo {volume} {83}},\ \bibinfo {pages} {245213} (\bibinfo
  {year} {2011})}\BibitemShut {NoStop}%
\bibitem [{\citenamefont {Radisavljevic}\ \emph {et~al.}(2011)\citenamefont
  {Radisavljevic}, \citenamefont {Radenovic}, \citenamefont {Brivio},
  \citenamefont {Giacometti},\ and\ \citenamefont {Kis}}]{Radisavljevic2011}%
  \BibitemOpen
  \bibfield  {author} {\bibinfo {author} {\bibfnamefont {B.}~\bibnamefont
  {Radisavljevic}}, \bibinfo {author} {\bibfnamefont {A.}~\bibnamefont
  {Radenovic}}, \bibinfo {author} {\bibfnamefont {J.}~\bibnamefont {Brivio}},
  \bibinfo {author} {\bibfnamefont {V.}~\bibnamefont {Giacometti}}, \ and\
  \bibinfo {author} {\bibfnamefont {A.}~\bibnamefont {Kis}},\ }\href@noop {}
  {\bibfield  {journal} {\bibinfo  {journal} {Nature Nanotechnology}\ }\textbf
  {\bibinfo {volume} {6}},\ \bibinfo {pages} {147} (\bibinfo {year}
  {2011})}\BibitemShut {NoStop}%
\bibitem [{\citenamefont {Son}\ \emph {et~al.}(2006{\natexlab{a}})\citenamefont
  {Son}, \citenamefont {Cohen}, ,\ and\ \citenamefont {Louie}}]{Son2006}%
  \BibitemOpen
  \bibfield  {author} {\bibinfo {author} {\bibfnamefont {Y.~W.}\ \bibnamefont
  {Son}}, \bibinfo {author} {\bibfnamefont {M.~L.}\ \bibnamefont {Cohen}}, , \
  and\ \bibinfo {author} {\bibfnamefont {S.~G.}\ \bibnamefont {Louie}},\
  }\href@noop {} {\bibfield  {journal} {\bibinfo  {journal} {Nature (London)}\
  }\textbf {\bibinfo {volume} {444}},\ \bibinfo {pages} {347} (\bibinfo {year}
  {2006}{\natexlab{a}})}\BibitemShut {NoStop}%
\bibitem [{\citenamefont {Yang}\ \emph
  {et~al.}(2007{\natexlab{a}})\citenamefont {Yang}, \citenamefont {Cohen},\
  and\ \citenamefont {Louie}}]{Yang2007}%
  \BibitemOpen
  \bibfield  {author} {\bibinfo {author} {\bibfnamefont {L.}~\bibnamefont
  {Yang}}, \bibinfo {author} {\bibfnamefont {M.~L.}\ \bibnamefont {Cohen}}, \
  and\ \bibinfo {author} {\bibfnamefont {S.~G.}\ \bibnamefont {Louie}},\
  }\href@noop {} {\bibfield  {journal} {\bibinfo  {journal} {Nano Letters}\
  }\textbf {\bibinfo {volume} {7}},\ \bibinfo {pages} {3112} (\bibinfo {year}
  {2007}{\natexlab{a}})}\BibitemShut {NoStop}%
\bibitem [{\citenamefont {Wang}\ \emph {et~al.}(2008)\citenamefont {Wang},
  \citenamefont {Ouyang}, \citenamefont {Li}, \citenamefont {Wang},
  \citenamefont {Guo},\ and\ \citenamefont {Dai}}]{Wang2008}%
  \BibitemOpen
  \bibfield  {author} {\bibinfo {author} {\bibfnamefont {X.}~\bibnamefont
  {Wang}}, \bibinfo {author} {\bibfnamefont {Y.}~\bibnamefont {Ouyang}},
  \bibinfo {author} {\bibfnamefont {X.}~\bibnamefont {Li}}, \bibinfo {author}
  {\bibfnamefont {H.}~\bibnamefont {Wang}}, \bibinfo {author} {\bibfnamefont
  {J.}~\bibnamefont {Guo}}, \ and\ \bibinfo {author} {\bibfnamefont
  {H.}~\bibnamefont {Dai}},\ }\href@noop {} {\bibfield  {journal} {\bibinfo
  {journal} {Phys. Rev. Lett}\ }\textbf {\bibinfo {volume} {100}},\ \bibinfo
  {pages} {206803} (\bibinfo {year} {2008})}\BibitemShut {NoStop}%
\bibitem [{\citenamefont {Li}\ \emph {et~al.}(2014)\citenamefont {Li},
  \citenamefont {Yu}, \citenamefont {Ye}, \citenamefont {Ge}, \citenamefont
  {Ou}, \citenamefont {Wu}, \citenamefont {Feng}, \citenamefont {Chen},\ and\
  \citenamefont {Zhang}}]{Li2014}%
  \BibitemOpen
  \bibfield  {author} {\bibinfo {author} {\bibfnamefont {L.}~\bibnamefont
  {Li}}, \bibinfo {author} {\bibfnamefont {Y.}~\bibnamefont {Yu}}, \bibinfo
  {author} {\bibfnamefont {G.~J.}\ \bibnamefont {Ye}}, \bibinfo {author}
  {\bibfnamefont {Q.}~\bibnamefont {Ge}}, \bibinfo {author} {\bibfnamefont
  {X.}~\bibnamefont {Ou}}, \bibinfo {author} {\bibfnamefont {H.}~\bibnamefont
  {Wu}}, \bibinfo {author} {\bibfnamefont {D.}~\bibnamefont {Feng}}, \bibinfo
  {author} {\bibfnamefont {X.~H.}\ \bibnamefont {Chen}}, \ and\ \bibinfo
  {author} {\bibfnamefont {Y.}~\bibnamefont {Zhang}},\ }\href@noop {}
  {\bibfield  {journal} {\bibinfo  {journal} {Nature Nanotechnology}\ }\textbf
  {\bibinfo {volume} {9}},\ \bibinfo {pages} {372} (\bibinfo {year}
  {2014})}\BibitemShut {NoStop}%
\bibitem [{\citenamefont {Liu}\ \emph {et~al.}(2014)\citenamefont {Liu},
  \citenamefont {Neal}, \citenamefont {Zhu}, \citenamefont {Xu}, \citenamefont
  {Tomanek},\ and\ \citenamefont {Ye}}]{Liu2014}%
  \BibitemOpen
  \bibfield  {author} {\bibinfo {author} {\bibfnamefont {H.}~\bibnamefont
  {Liu}}, \bibinfo {author} {\bibfnamefont {A.~T.}\ \bibnamefont {Neal}},
  \bibinfo {author} {\bibfnamefont {Z.}~\bibnamefont {Zhu}}, \bibinfo {author}
  {\bibfnamefont {X.}~\bibnamefont {Xu}}, \bibinfo {author} {\bibfnamefont
  {D.}~\bibnamefont {Tomanek}}, \ and\ \bibinfo {author} {\bibfnamefont
  {P.~D.}\ \bibnamefont {Ye}},\ }\href@noop {} {\bibfield  {journal} {\bibinfo
  {journal} {ACS Nano}\ }\textbf {\bibinfo {volume} {8}},\ \bibinfo {pages}
  {4033} (\bibinfo {year} {2014})}\BibitemShut {NoStop}%
\bibitem [{\citenamefont {Xia}\ \emph {et~al.}(2014)\citenamefont {Xia},
  \citenamefont {Wang},\ and\ \citenamefont {Jia}}]{Xia2014}%
  \BibitemOpen
  \bibfield  {author} {\bibinfo {author} {\bibfnamefont {F.}~\bibnamefont
  {Xia}}, \bibinfo {author} {\bibfnamefont {H.}~\bibnamefont {Wang}}, \ and\
  \bibinfo {author} {\bibfnamefont {Y.}~\bibnamefont {Jia}},\ }\href@noop {}
  {\bibfield  {journal} {\bibinfo  {journal} {Nature Communications}\ }\textbf
  {\bibinfo {volume} {5}},\ \bibinfo {pages} {4458} (\bibinfo {year}
  {2014})}\BibitemShut {NoStop}%
\bibitem [{\citenamefont {Castellanos-Gomez}\ \emph {et~al.}(2014)\citenamefont
  {Castellanos-Gomez}, \citenamefont {Vicarelli}, \citenamefont {Prada},
  \citenamefont {Island}, \citenamefont {Narasimha-Acharya}, \citenamefont
  {Blanter}, \citenamefont {Groenendijk}, \citenamefont {Buscema},
  \citenamefont {Steele}, \citenamefont {Alvarez}, \citenamefont {Zandbergen},
  \citenamefont {Palacios},\ and\ \citenamefont {van~der Zant}}]{Gomez2014}%
  \BibitemOpen
  \bibfield  {author} {\bibinfo {author} {\bibfnamefont {A.}~\bibnamefont
  {Castellanos-Gomez}}, \bibinfo {author} {\bibfnamefont {L.}~\bibnamefont
  {Vicarelli}}, \bibinfo {author} {\bibfnamefont {E.}~\bibnamefont {Prada}},
  \bibinfo {author} {\bibfnamefont {J.~O.}\ \bibnamefont {Island}}, \bibinfo
  {author} {\bibfnamefont {K.~L.}\ \bibnamefont {Narasimha-Acharya}}, \bibinfo
  {author} {\bibfnamefont {S.~I.}\ \bibnamefont {Blanter}}, \bibinfo {author}
  {\bibfnamefont {D.~J.}\ \bibnamefont {Groenendijk}}, \bibinfo {author}
  {\bibfnamefont {M.}~\bibnamefont {Buscema}}, \bibinfo {author} {\bibfnamefont
  {G.~A.}\ \bibnamefont {Steele}}, \bibinfo {author} {\bibfnamefont {J.~V.}\
  \bibnamefont {Alvarez}}, \bibinfo {author} {\bibfnamefont {H.~W.}\
  \bibnamefont {Zandbergen}}, \bibinfo {author} {\bibfnamefont {J.~J.}\
  \bibnamefont {Palacios}}, \ and\ \bibinfo {author} {\bibfnamefont {H.~S.~J.}\
  \bibnamefont {van~der Zant}},\ }\href@noop {} {\bibfield  {journal} {\bibinfo
   {journal} {2D Materials}\ }\textbf {\bibinfo {volume} {1}},\ \bibinfo
  {pages} {025001} (\bibinfo {year} {2014})}\BibitemShut {NoStop}%
\bibitem [{\citenamefont {Koenig}\ \emph {et~al.}(2014)\citenamefont {Koenig},
  \citenamefont {Doganov}, \citenamefont {Schmidt}, \citenamefont {Neto},\ and\
  \citenamefont {Oezyilmaz}}]{Koenig2014}%
  \BibitemOpen
  \bibfield  {author} {\bibinfo {author} {\bibfnamefont {S.~P.}\ \bibnamefont
  {Koenig}}, \bibinfo {author} {\bibfnamefont {R.~A.}\ \bibnamefont {Doganov}},
  \bibinfo {author} {\bibfnamefont {H.}~\bibnamefont {Schmidt}}, \bibinfo
  {author} {\bibfnamefont {A.~H.~C.}\ \bibnamefont {Neto}}, \ and\ \bibinfo
  {author} {\bibfnamefont {B.}~\bibnamefont {Oezyilmaz}},\ }\href@noop {}
  {\bibfield  {journal} {\bibinfo  {journal} {Appl. Phys. Lett.}\ }\textbf
  {\bibinfo {volume} {104}},\ \bibinfo {pages} {103106} (\bibinfo {year}
  {2014})}\BibitemShut {NoStop}%
\bibitem [{\citenamefont {Morita}(1986)}]{Morita1986}%
  \BibitemOpen
  \bibfield  {author} {\bibinfo {author} {\bibfnamefont {A.}~\bibnamefont
  {Morita}},\ }\href@noop {} {\bibfield  {journal} {\bibinfo  {journal} {Appl.
  Phys. A}\ }\textbf {\bibinfo {volume} {39}},\ \bibinfo {pages} {227}
  (\bibinfo {year} {1986})}\BibitemShut {NoStop}%
\bibitem [{\citenamefont {Warschauer}(1963)}]{Warschauer1963}%
  \BibitemOpen
  \bibfield  {author} {\bibinfo {author} {\bibfnamefont {D.}~\bibnamefont
  {Warschauer}},\ }\href@noop {} {\bibfield  {journal} {\bibinfo  {journal} {J.
  Appl. Phys.}\ }\textbf {\bibinfo {volume} {34}},\ \bibinfo {pages} {1853}
  (\bibinfo {year} {1963})}\BibitemShut {NoStop}%
\bibitem [{\citenamefont {Narita}\ \emph {et~al.}(1983)\citenamefont {Narita},
  \citenamefont {Akahama}, \citenamefont {Tsukiyamaa}, \citenamefont {Muroa},
  \citenamefont {Moria}, \citenamefont {Endo}, \citenamefont {Taniguchi},
  \citenamefont {Seki}, \citenamefont {Suga}, \citenamefont {Mikuni},\ and\
  \citenamefont {Kanzaki}}]{Narita1983}%
  \BibitemOpen
  \bibfield  {author} {\bibinfo {author} {\bibfnamefont {S.}~\bibnamefont
  {Narita}}, \bibinfo {author} {\bibfnamefont {Y.}~\bibnamefont {Akahama}},
  \bibinfo {author} {\bibfnamefont {Y.}~\bibnamefont {Tsukiyamaa}}, \bibinfo
  {author} {\bibfnamefont {K.}~\bibnamefont {Muroa}}, \bibinfo {author}
  {\bibfnamefont {S.}~\bibnamefont {Moria}}, \bibinfo {author} {\bibfnamefont
  {S.}~\bibnamefont {Endo}}, \bibinfo {author} {\bibfnamefont {M.}~\bibnamefont
  {Taniguchi}}, \bibinfo {author} {\bibfnamefont {M.}~\bibnamefont {Seki}},
  \bibinfo {author} {\bibfnamefont {S.}~\bibnamefont {Suga}}, \bibinfo {author}
  {\bibfnamefont {A.}~\bibnamefont {Mikuni}}, \ and\ \bibinfo {author}
  {\bibfnamefont {H.}~\bibnamefont {Kanzaki}},\ }\href@noop {} {\bibfield
  {journal} {\bibinfo  {journal} {Physica B+C}\ }\textbf {\bibinfo {volume}
  {117}},\ \bibinfo {pages} {422} (\bibinfo {year} {1983})}\BibitemShut
  {NoStop}%
\bibitem [{\citenamefont {Maruyama}\ \emph {et~al.}(1981)\citenamefont
  {Maruyama}, \citenamefont {Suzuki}, \citenamefont {Kobayashi},\ and\
  \citenamefont {Tanuma}}]{Maruyama1981}%
  \BibitemOpen
  \bibfield  {author} {\bibinfo {author} {\bibfnamefont {Y.}~\bibnamefont
  {Maruyama}}, \bibinfo {author} {\bibfnamefont {S.}~\bibnamefont {Suzuki}},
  \bibinfo {author} {\bibfnamefont {K.}~\bibnamefont {Kobayashi}}, \ and\
  \bibinfo {author} {\bibfnamefont {S.}~\bibnamefont {Tanuma}},\ }\href@noop {}
  {\bibfield  {journal} {\bibinfo  {journal} {Physica B+C}\ }\textbf {\bibinfo
  {volume} {105}},\ \bibinfo {pages} {99} (\bibinfo {year} {1981})}\BibitemShut
  {NoStop}%
\bibitem [{\citenamefont {Liang}\ \emph {et~al.}(2014)\citenamefont {Liang},
  \citenamefont {Wang}, \citenamefont {Lin}, \citenamefont {Sumpter},
  \citenamefont {Meunier},\ and\ \citenamefont {Pan}}]{Liang2014}%
  \BibitemOpen
  \bibfield  {author} {\bibinfo {author} {\bibfnamefont {L.}~\bibnamefont
  {Liang}}, \bibinfo {author} {\bibfnamefont {J.}~\bibnamefont {Wang}},
  \bibinfo {author} {\bibfnamefont {W.}~\bibnamefont {Lin}}, \bibinfo {author}
  {\bibfnamefont {B.~G.}\ \bibnamefont {Sumpter}}, \bibinfo {author}
  {\bibfnamefont {V.}~\bibnamefont {Meunier}}, \ and\ \bibinfo {author}
  {\bibfnamefont {M.}~\bibnamefont {Pan}},\ }\href@noop {} {\bibfield
  {journal} {\bibinfo  {journal} {Nano Lett.}\ }\textbf {\bibinfo {volume}
  {14}},\ \bibinfo {pages} {6400} (\bibinfo {year} {2014})}\BibitemShut
  {NoStop}%
\bibitem [{\citenamefont {Tran}\ \emph {et~al.}(2014)\citenamefont {Tran},
  \citenamefont {Soklaski}, \citenamefont {Liang},\ and\ \citenamefont
  {Yang}}]{Tran20141}%
  \BibitemOpen
  \bibfield  {author} {\bibinfo {author} {\bibfnamefont {V.}~\bibnamefont
  {Tran}}, \bibinfo {author} {\bibfnamefont {R.}~\bibnamefont {Soklaski}},
  \bibinfo {author} {\bibfnamefont {Y.}~\bibnamefont {Liang}}, \ and\ \bibinfo
  {author} {\bibfnamefont {L.}~\bibnamefont {Yang}},\ }\href@noop {} {\bibfield
   {journal} {\bibinfo  {journal} {Phys. Rev. B}\ }\textbf {\bibinfo {volume}
  {89}},\ \bibinfo {pages} {235319} (\bibinfo {year} {2014})}\BibitemShut
  {NoStop}%
\bibitem [{\citenamefont {Qiao}\ \emph {et~al.}(2014)\citenamefont {Qiao},
  \citenamefont {Kong}, \citenamefont {Hu}, \citenamefont {Yang},\ and\
  \citenamefont {Ji}}]{Qiao2014}%
  \BibitemOpen
  \bibfield  {author} {\bibinfo {author} {\bibfnamefont {J.}~\bibnamefont
  {Qiao}}, \bibinfo {author} {\bibfnamefont {X.}~\bibnamefont {Kong}}, \bibinfo
  {author} {\bibfnamefont {Z.-X.}\ \bibnamefont {Hu}}, \bibinfo {author}
  {\bibfnamefont {F.}~\bibnamefont {Yang}}, \ and\ \bibinfo {author}
  {\bibfnamefont {W.}~\bibnamefont {Ji}},\ }\href@noop {} {\bibfield  {journal}
  {\bibinfo  {journal} {Nature Communications}\ }\textbf {\bibinfo {volume}
  {5}},\ \bibinfo {pages} {4475} (\bibinfo {year} {2014})}\BibitemShut
  {NoStop}%
\bibitem [{\citenamefont {Wei}\ and\ \citenamefont {Peng}(2014)}]{Wei2014}%
  \BibitemOpen
  \bibfield  {author} {\bibinfo {author} {\bibfnamefont {Q.}~\bibnamefont
  {Wei}}\ and\ \bibinfo {author} {\bibfnamefont {X.}~\bibnamefont {Peng}},\
  }\href@noop {} {\bibfield  {journal} {\bibinfo  {journal} {Appl. Phys. Lett}\
  }\textbf {\bibinfo {volume} {104}},\ \bibinfo {pages} {251915} (\bibinfo
  {year} {2014})}\BibitemShut {NoStop}%
\bibitem [{\citenamefont {Zhang}\ \emph {et~al.}(2014)\citenamefont {Zhang},
  \citenamefont {Liu}, \citenamefont {Cheng}, \citenamefont {Wei},
  \citenamefont {Liang}, \citenamefont {Fan}, \citenamefont {Shi},
  \citenamefont {Tang},\ and\ \citenamefont {Zhang}}]{Zhang2014}%
  \BibitemOpen
  \bibfield  {author} {\bibinfo {author} {\bibfnamefont {J.}~\bibnamefont
  {Zhang}}, \bibinfo {author} {\bibfnamefont {H.}~\bibnamefont {Liu}}, \bibinfo
  {author} {\bibfnamefont {L.}~\bibnamefont {Cheng}}, \bibinfo {author}
  {\bibfnamefont {J.}~\bibnamefont {Wei}}, \bibinfo {author} {\bibfnamefont
  {J.}~\bibnamefont {Liang}}, \bibinfo {author} {\bibfnamefont
  {D.}~\bibnamefont {Fan}}, \bibinfo {author} {\bibfnamefont {J.}~\bibnamefont
  {Shi}}, \bibinfo {author} {\bibfnamefont {X.}~\bibnamefont {Tang}}, \ and\
  \bibinfo {author} {\bibfnamefont {Q.~J.}\ \bibnamefont {Zhang}},\ }\href@noop
  {} {\bibfield  {journal} {\bibinfo  {journal} {cond-mat/arXiv:1405.3348}\ }
  (\bibinfo {year} {2014})}\BibitemShut {NoStop}%
\bibitem [{\citenamefont {Lv}\ \emph {et~al.}(2014)\citenamefont {Lv},
  \citenamefont {Lu}, \citenamefont {Shao},\ and\ \citenamefont
  {Sun}}]{Lv2014}%
  \BibitemOpen
  \bibfield  {author} {\bibinfo {author} {\bibfnamefont {H.~Y.}\ \bibnamefont
  {Lv}}, \bibinfo {author} {\bibfnamefont {W.~J.}\ \bibnamefont {Lu}}, \bibinfo
  {author} {\bibfnamefont {D.~F.}\ \bibnamefont {Shao}}, \ and\ \bibinfo
  {author} {\bibfnamefont {Y.~P.}\ \bibnamefont {Sun}},\ }\href@noop {}
  {\bibfield  {journal} {\bibinfo  {journal} {cond-mat/arXiv:1404.5171}\ }
  (\bibinfo {year} {2014})}\BibitemShut {NoStop}%
\bibitem [{\citenamefont {Gong}\ \emph {et~al.}(2014)\citenamefont {Gong},
  \citenamefont {Zhang}, \citenamefont {Ji},\ and\ \citenamefont
  {Guo}}]{Gong2014}%
  \BibitemOpen
  \bibfield  {author} {\bibinfo {author} {\bibfnamefont {K.}~\bibnamefont
  {Gong}}, \bibinfo {author} {\bibfnamefont {L.}~\bibnamefont {Zhang}},
  \bibinfo {author} {\bibfnamefont {W.}~\bibnamefont {Ji}}, \ and\ \bibinfo
  {author} {\bibfnamefont {H.}~\bibnamefont {Guo}},\ }\href@noop {} {\bibfield
  {journal} {\bibinfo  {journal} {cond-mat/arXiv:1404.7207}\ } (\bibinfo {year}
  {2014})}\BibitemShut {NoStop}%
\bibitem [{\citenamefont {Rudenko}\ and\ \citenamefont
  {Katsnelson}(2014)}]{Rudenko2014}%
  \BibitemOpen
  \bibfield  {author} {\bibinfo {author} {\bibfnamefont {A.~N.}\ \bibnamefont
  {Rudenko}}\ and\ \bibinfo {author} {\bibfnamefont {M.~I.}\ \bibnamefont
  {Katsnelson}},\ }\href@noop {} {\bibfield  {journal} {\bibinfo  {journal}
  {Phys. Rev. B}\ }\textbf {\bibinfo {volume} {89}},\ \bibinfo {pages} {201408}
  (\bibinfo {year} {2014})}\BibitemShut {NoStop}%
\bibitem [{\citenamefont {Tran}\ and\ \citenamefont {Yang}(2014)}]{Tran2014}%
  \BibitemOpen
  \bibfield  {author} {\bibinfo {author} {\bibfnamefont {V.}~\bibnamefont
  {Tran}}\ and\ \bibinfo {author} {\bibfnamefont {L.}~\bibnamefont {Yang}},\
  }\href@noop {} {\bibfield  {journal} {\bibinfo  {journal} {Phys. Rev. B}\
  }\textbf {\bibinfo {volume} {89}},\ \bibinfo {pages} {245407} (\bibinfo
  {year} {2014})}\BibitemShut {NoStop}%
\bibitem [{\citenamefont {Dolui}\ \emph {et~al.}(2012)\citenamefont {Dolui},
  \citenamefont {Pemmaraj},\ and\ \citenamefont {Sanvito}}]{Dolui2012}%
  \BibitemOpen
  \bibfield  {author} {\bibinfo {author} {\bibfnamefont {K.}~\bibnamefont
  {Dolui}}, \bibinfo {author} {\bibfnamefont {C.~D.}\ \bibnamefont {Pemmaraj}},
  \ and\ \bibinfo {author} {\bibfnamefont {S.}~\bibnamefont {Sanvito}},\
  }\href@noop {} {\bibfield  {journal} {\bibinfo  {journal} {ACS Nano}\
  }\textbf {\bibinfo {volume} {6}},\ \bibinfo {pages} {4823} (\bibinfo {year}
  {2012})}\BibitemShut {NoStop}%
\bibitem [{\citenamefont {Son}\ \emph {et~al.}(2006{\natexlab{b}})\citenamefont
  {Son}, \citenamefont {Cohen},\ and\ \citenamefont {Louie}}]{Son20061}%
  \BibitemOpen
  \bibfield  {author} {\bibinfo {author} {\bibfnamefont {Y.-W.}\ \bibnamefont
  {Son}}, \bibinfo {author} {\bibfnamefont {M.~L.}\ \bibnamefont {Cohen}}, \
  and\ \bibinfo {author} {\bibfnamefont {S.~G.}\ \bibnamefont {Louie}},\
  }\href@noop {} {\bibfield  {journal} {\bibinfo  {journal} {Phys. Rev. Lett.}\
  }\textbf {\bibinfo {volume} {97}},\ \bibinfo {pages} {216803} (\bibinfo
  {year} {2006}{\natexlab{b}})}\BibitemShut {NoStop}%
\bibitem [{\citenamefont {Zhao}\ \emph {et~al.}(2004)\citenamefont {Zhao},
  \citenamefont {Wei}, \citenamefont {Yang},\ and\ \citenamefont
  {Chou}}]{Zhao2004}%
  \BibitemOpen
  \bibfield  {author} {\bibinfo {author} {\bibfnamefont {X.~Y.}\ \bibnamefont
  {Zhao}}, \bibinfo {author} {\bibfnamefont {C.~M.}\ \bibnamefont {Wei}},
  \bibinfo {author} {\bibfnamefont {L.}~\bibnamefont {Yang}}, \ and\ \bibinfo
  {author} {\bibfnamefont {M.~Y.}\ \bibnamefont {Chou}},\ }\href@noop {}
  {\bibfield  {journal} {\bibinfo  {journal} {Phys. Rev. Lett.}\ }\textbf
  {\bibinfo {volume} {92}},\ \bibinfo {pages} {236805} (\bibinfo {year}
  {2004})}\BibitemShut {NoStop}%
\bibitem [{\citenamefont {Yang}\ \emph
  {et~al.}(2007{\natexlab{b}})\citenamefont {Yang}, \citenamefont {Park},
  \citenamefont {Son}, \citenamefont {Cohen},\ and\ \citenamefont
  {Louie}}]{Yang20071}%
  \BibitemOpen
  \bibfield  {author} {\bibinfo {author} {\bibfnamefont {L.}~\bibnamefont
  {Yang}}, \bibinfo {author} {\bibfnamefont {C.-H.}\ \bibnamefont {Park}},
  \bibinfo {author} {\bibfnamefont {Y.-W.}\ \bibnamefont {Son}}, \bibinfo
  {author} {\bibfnamefont {M.~L.}\ \bibnamefont {Cohen}}, \ and\ \bibinfo
  {author} {\bibfnamefont {S.~G.}\ \bibnamefont {Louie}},\ }\href@noop {}
  {\bibfield  {journal} {\bibinfo  {journal} {Phys. Rev. Lett.}\ }\textbf
  {\bibinfo {volume} {99}},\ \bibinfo {pages} {186801} (\bibinfo {year}
  {2007}{\natexlab{b}})}\BibitemShut {NoStop}%
\bibitem [{\citenamefont {Park}\ and\ \citenamefont {Louie}(2008)}]{Park2008}%
  \BibitemOpen
  \bibfield  {author} {\bibinfo {author} {\bibfnamefont {C.-H.}\ \bibnamefont
  {Park}}\ and\ \bibinfo {author} {\bibfnamefont {S.~G.}\ \bibnamefont
  {Louie}},\ }\href@noop {} {\bibfield  {journal} {\bibinfo  {journal} {Nano
  Lett.}\ }\textbf {\bibinfo {volume} {8}},\ \bibinfo {pages} {2200} (\bibinfo
  {year} {2008})}\BibitemShut {NoStop}%
\bibitem [{\citenamefont {Zhang}\ and\ \citenamefont {Guo}(2008)}]{Zhang2008}%
  \BibitemOpen
  \bibfield  {author} {\bibinfo {author} {\bibfnamefont {Z.}~\bibnamefont
  {Zhang}}\ and\ \bibinfo {author} {\bibfnamefont {W.}~\bibnamefont {Guo}},\
  }\href@noop {} {\bibfield  {journal} {\bibinfo  {journal} {Phys. Rev. B}\
  }\textbf {\bibinfo {volume} {77}},\ \bibinfo {pages} {075403} (\bibinfo
  {year} {2008})}\BibitemShut {NoStop}%
\bibitem [{\citenamefont {Niu}\ \emph {et~al.}(2014)\citenamefont {Niu},
  \citenamefont {Yang}, \citenamefont {Si},\ and\ \citenamefont
  {Xue}}]{Niu2014}%
  \BibitemOpen
  \bibfield  {author} {\bibinfo {author} {\bibfnamefont {X.~N.}\ \bibnamefont
  {Niu}}, \bibinfo {author} {\bibfnamefont {D.~Z.}\ \bibnamefont {Yang}},
  \bibinfo {author} {\bibfnamefont {M.~S.}\ \bibnamefont {Si}}, \ and\ \bibinfo
  {author} {\bibfnamefont {D.~S.}\ \bibnamefont {Xue}},\ }\href@noop {}
  {\bibfield  {journal} {\bibinfo  {journal} {J. Applied Physics}\ }\textbf
  {\bibinfo {volume} {115}},\ \bibinfo {pages} {143706} (\bibinfo {year}
  {2014})}\BibitemShut {NoStop}%
\bibitem [{\citenamefont {Q~Wu}\ \emph {et~al.}(2014)\citenamefont {Q~Wu},
  \citenamefont {Yang}, \citenamefont {Huang},\ and\ \citenamefont
  {Feng}}]{Wu2014}%
  \BibitemOpen
  \bibfield  {author} {\bibinfo {author} {\bibfnamefont {L.~S.}\ \bibnamefont
  {Q~Wu}}, \bibinfo {author} {\bibfnamefont {M.}~\bibnamefont {Yang}}, \bibinfo
  {author} {\bibfnamefont {Z.}~\bibnamefont {Huang}}, \ and\ \bibinfo {author}
  {\bibfnamefont {Y.~P.}\ \bibnamefont {Feng}},\ }\href@noop {} {\bibfield
  {journal} {\bibinfo  {journal} {cond-mat/arxiv.org:1405.3077}\ } (\bibinfo
  {year} {2014})}\BibitemShut {NoStop}%
\bibitem [{\citenamefont {\'{O}Keeffe}\ \emph {et~al.}(2002)\citenamefont
  {\'{O}Keeffe}, \citenamefont {Wei},\ and\ \citenamefont {Cho}}]{Okeeffe2002}%
  \BibitemOpen
  \bibfield  {author} {\bibinfo {author} {\bibfnamefont {J.}~\bibnamefont
  {\'{O}Keeffe}}, \bibinfo {author} {\bibfnamefont {C.~Y.}\ \bibnamefont
  {Wei}}, \ and\ \bibinfo {author} {\bibfnamefont {K.~J.}\ \bibnamefont
  {Cho}},\ }\href@noop {} {\bibfield  {journal} {\bibinfo  {journal} {Appl.
  Phys. Lett.}\ }\textbf {\bibinfo {volume} {80}},\ \bibinfo {pages} {676}
  (\bibinfo {year} {2002})}\BibitemShut {NoStop}%
\bibitem [{\citenamefont {Barone}\ and\ \citenamefont
  {Peralta}(2008)}]{Barone2008}%
  \BibitemOpen
  \bibfield  {author} {\bibinfo {author} {\bibfnamefont {V.}~\bibnamefont
  {Barone}}\ and\ \bibinfo {author} {\bibfnamefont {J.~E.}\ \bibnamefont
  {Peralta}},\ }\href@noop {} {\bibfield  {journal} {\bibinfo  {journal} {Nano
  Letters}\ }\textbf {\bibinfo {volume} {8}},\ \bibinfo {pages} {2210}
  (\bibinfo {year} {2008})}\BibitemShut {NoStop}%
\bibitem [{\citenamefont {Ezawa}(2014)}]{Ezawa2014}%
  \BibitemOpen
  \bibfield  {author} {\bibinfo {author} {\bibfnamefont {M.}~\bibnamefont
  {Ezawa}},\ }\href@noop {} {\bibfield  {journal} {\bibinfo  {journal} {New J.
  Phys.}\ }\textbf {\bibinfo {volume} {16}},\ \bibinfo {pages} {115004}
  (\bibinfo {year} {2014})}\BibitemShut {NoStop}%
\bibitem [{\citenamefont {Datta}(1995)}]{Datta1995}%
  \BibitemOpen
  \bibfield  {author} {\bibinfo {author} {\bibfnamefont {S.}~\bibnamefont
  {Datta}},\ }\href@noop {} {\emph {\bibinfo {title} {Electronic Transport in
  Mesoscopic Systems.}}}\ (\bibinfo  {publisher} {Cambridge University Press,
  Cambridge, England},\ \bibinfo {year} {1995})\BibitemShut {NoStop}%
\bibitem [{\citenamefont {Datta}(2005)}]{Datta2005}%
  \BibitemOpen
  \bibfield  {author} {\bibinfo {author} {\bibfnamefont {S.}~\bibnamefont
  {Datta}},\ }\href@noop {} {\emph {\bibinfo {title} {Quantum Transport: Atom
  to Transistor.}}}\ (\bibinfo  {publisher} {Cambridge University Press,
  Cambridge, England},\ \bibinfo {year} {2005})\BibitemShut {NoStop}%
\end{thebibliography}%

\end{document}